\documentclass[aps,prx,twocolumn,notitlepage,amsfonts,citeautoscript,superscriptaddress,showpacs]{revtex4-1}

\usepackage{natbib}
\usepackage{listings}
\usepackage{color}

\usepackage{amsmath,amssymb,mathrsfs}
\usepackage{latexsym}
\usepackage{graphicx} 
\usepackage{grffile}
\usepackage{epstopdf}
\usepackage{graphicx,epstopdf,color}
\usepackage{amsfonts}
\usepackage{amsmath,amssymb,mathrsfs}
\usepackage{hyperref}
\usepackage[dvipsnames]{xcolor}
\usepackage{url}
\usepackage{soul}
\usepackage{cancel}
\usepackage{physics}

\usepackage{cleveref}
\crefname{equation}{Eq.}{Eqs.}
\Crefname{equation}{Equation}{Equations}
\crefname{figure}{Fig.}{Figs.}
\Crefname{figure}{Figure}{Figures}
\crefname{section}{Sec.}{Secs.}
\Crefname{section}{Section}{Sections}
\crefname{appendix}{Appendix}{Apps.}
\Crefname{appendix}{Appendix}{Apps.}
\crefname{paragraph}{Sec.}{Secs.}
\crefname{table}{Table}{Tables}

\usepackage{xr}

\graphicspath{{nb_fig/}}

\usepackage{epstopdf}

\newcommand{\phiext}{\varphi_\mathrm{ext}}

\newcommand{\LPENS}{\affiliation{Laboratoire de Physique de l'Ecole Normale Supérieure, ENS-PSL, CNRS, Sorbonne Université, Université Paris Cité, Centre Automatique et Systèmes, Mines Paris, Université PSL, Inria, Paris, France}}

\newcommand{\LPTHE}{\affiliation{Laboratoire de Physique Th\'{e}orique et Hautes Energies, Sorbonne Universit\'{e} and CNRS UMR 7589, 4 place Jussieu, 75252 Paris Cedex 05, France}}

\begin{document}
\title{Symmetries and model reduction in a $\cos 2\varphi$ transmon qubit}

\author{E.~Roverc'h}
\LPENS
\author{K.~ Gerashchenko}
\LPENS
\author{A.~Borgognoni}
\LPENS
\author{B.~Dou\c{c}ot}
\LPTHE
\author{A.~Petrescu}
\LPENS
\author{R.~Robin}
\LPENS
\author{Z.~Leghtas}
\email[]{zaki.leghtas@phys.ens.fr}
\LPENS
\date{\today}

\begin{abstract}
Engineered symmetries in superconducting circuits can protect quantum information; much like selection rules in atomic physics, they forbid transitions between the qubit states. Protected circuits, however, combine several loops and islands and host multiple nonlinearly coupled modes, calling for reduced low-energy models that inherit the circuit symmetries and make the protection manifest. We develop such a reduction for the $\cos 2\varphi$ transmon qubit, recently realized experimentally with a two-Cooper-pair interferometer, whose circuit hosts three coupled modes. Exploiting the separation of energy scales between the heavy transmon mode and the light internal modes of the interferometer, we derive an effective single-band Born–Oppenheimer model, including its geometric correction, and extend it to several bands. We identify the symmetries of the full and reduced Hamiltonians and derive exact selection rules. Finally, we identify a parameter regime beyond the single-band description that suppresses flux-noise dephasing without degrading charge-noise protection.
\end{abstract}
\maketitle

\section{Introduction}
In the ongoing quest for longer coherence times in superconducting circuits, protected qubits are promising candidates. To build such a qubit, one assembles a complex arrangement of circuit elements designed to embed symmetries directly into the Hamiltonian. One then identifies two energy levels, separated by a spectral gap from the rest of the spectrum, and residing in different parity sectors of the underlying symmetry. This symmetry enforces selection rules that forbid transitions between the qubit states, protecting it from dominant decoherence channels. Notable examples include the $0-\pi$ \cite{Gyenis0pi, Groszkowski2018}, and $\cos 2\varphi$ qubits \cite{Doucot2002,Ioffe2002,Gladchenko2008,Bell2014,Smith2020, Larsen2020, Dodge2023, Messelot2024, Brosco2024flowermon, Hays2025, Roverch2026, Zhurbina2026, Sanchez2026}, which display Cooper-pair-parity symmetry; the bifluxon \cite{Kalashnikov2020} and integer fluxonium \cite{Ardati2024, Mencia2024} qubits, which display fluxon-parity symmetry; and Gottesman-Kitaev-Preskill (GKP) qubits \cite{GKP2001, Rymarz2021, Nguyen2025}, which combine two translational symmetries in $IQ$ space. These circuits are composed of multiple elements — inductors, capacitors, Josephson junctions (JJs), and sometimes more exotic elements — forming loops and islands, and housing several nonlinearly coupled modes. This complexity obscures circuit design, which requires navigating the parameter space of circuit elements in order to maximize qubit protection. An enlightening solution is to derive effective low-energy models exploiting a separation of energy scales, elucidating the relationship between qubit protection and circuit parameters.

\begin{figure*}
    \centering
    \includegraphics[scale=1]{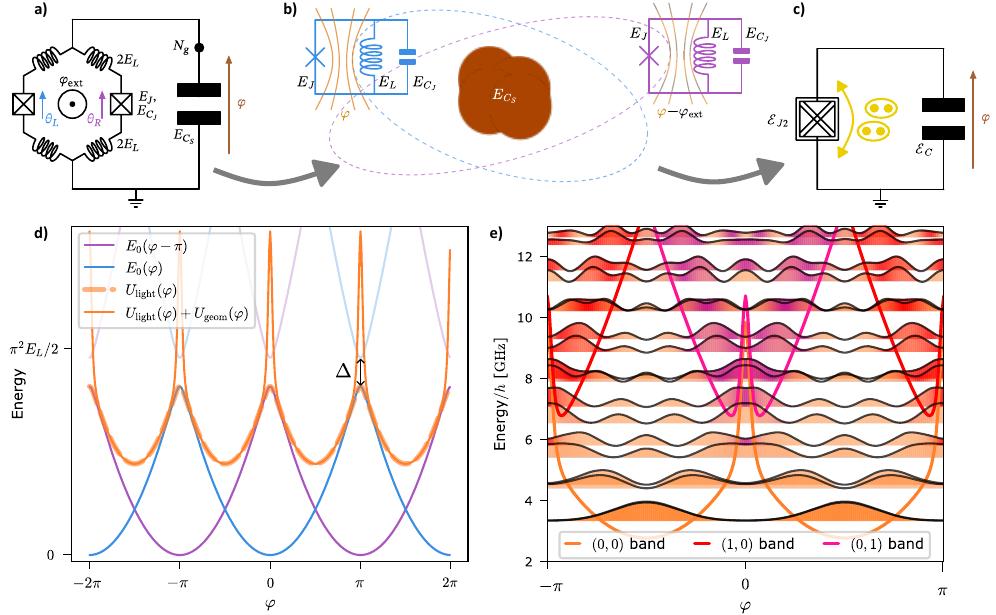}
    \caption{{\bf $\cos2\varphi$ transmon circuit and the Born-Oppenheimer approximation.} (a) Lumped element diagram of our circuit (see text).
    (b) Atomic analogy for Hamiltonian \eqref{eq:Hfull_to_Hlight}. A heavy nucleus ($\varphi$ mode) binds two light electrons ($\theta_{L/R}$ fluxonium modes). The usual Coulomb interactions are replaced by a $\varphi$-dependent flux bias for each fluxonium ($\varphi$ and $\varphi - \phiext$ respectively).
    (c) Effective single-mode circuit obtained from the single-band BO approximation.
    (d) Potential energy (y-axis) versus $\varphi$ (x-axis) for the single-band BO model at $\phiext = \pi$.
    The light-mode potential $U_\mathrm{light}$ (dashed orange line) is the sum of the ground-state energies of the two fluxonium circuits (blue and purple lines) that are separated by a gap $\Delta$ from their excited states (faint blue and purple lines). A significant correction from the geometric potential $U_\mathrm{geom}$ must be included to obtain the lowest BO band (orange line).
    (e) Potential energy (y-axis) of the first 3 bands (solid lines) versus $\varphi$ (x-axis), at $\phiext=\pi$. The squared Euclidean norm of the multi-component eigenstates $\lVert \overrightarrow{\Phi} \rVert^2 (\varphi)$ are overlaid, and offset by their corresponding eigenvalue. The color indicates the relative weight of each component in the eigenstate.
    }
    \label{fig:fig1}
\end{figure*}

We address this challenge for a $\cos 2\varphi$ qubit, conceptually consisting of a superconducting island of charging energy $\mathcal{E}_C$ shunted by a Cooper-quartet tunneling element of energy $\mathcal{E}_{J2}$. At offset charge $\mathcal{N}_g$, its Hamiltonian reads
\begin{equation}
    \mathcal{H}_{\mathrm{cos}2\varphi}=4\mathcal{E}_C (\mathcal{\hat N}-\mathcal{N}_g)^2+\mathcal{E}_{J2}\cos(2\hat \varphi)\;,
    \label{eq:Hcos2phi}
\end{equation}
where $\hat \varphi$ is the phase variable that lives on a circle \cite{Devoret2021DoesBJ} and is conjugate to charge $\mathcal{\hat N}$. More specifically, we consider the $\cos 2\varphi$ qubit implementation based on a Kinetic Interference coTunneling Element (KITE) \cite{Smith2020,Smith2022, Smith2025}, and recently experimentally realized in the soft transmon regime \cite{Roverch2026}. It consists of two identical arms enclosing half a flux quantum, shunted by a large capacitance of energy $E_{C_S}$. Each arm is composed of a small Josephson junction of Josephson and capacitive energies $(E_J, E_{C_J})$ in series with an inductor of energy $E_L$ (Fig. \ref{fig:fig1}a). The resulting circuit houses three non-linearly coupled modes. The questions that arise are then: how can this circuit be mapped to a $\cos 2\varphi$ qubit? What are the Hamiltonian symmetries and how do they translate into selection rules? How robust are these selection rules to symmetry breaking errors such as deviations from half a flux quantum, or disorder in the interferometer arms? And finally, which parameter regime for $(E_J, E_{C_J}, E_L, E_{C_S})$ maximizes protection against main sources of noise, namely: charge noise on the island, flux noise in the loop and dielectric loss?

In this paper, we perform a Born-Oppenheimer (BO) approximation \cite{BO1927} controlled 
by the small capacitive ratio $\epsilon = E_{C_S}/E_{C_J} \ll 1$. We first 
derive an effective single-band model retaining all first-order terms in 
$\epsilon$, including a geometric correction to the potential. 
This single-band model exhibits a $\pi$-periodic potential, making the 
$\cos 2\varphi$ nature of the qubit manifest. We then extend this analysis 
to higher order in $\epsilon$ by including higher bands of the light modes 
--- in direct analogy with a diatomic molecule \cite{Morse1929}, where nuclear vibrational 
states residing on different electronic potential energy surfaces give rise 
to corrections beyond the leading adiabatic approximation \cite{Orlikowski1974}. We then analyze 
the symmetries of the full three-mode and reduced single-band Hamiltonians, 
and derive selection rules for all charge and flux branch operators. This 
analysis provides a rigorous account of the experimentally observed 
limitation of this qubit by flux noise. Finally, we identify a pathway in 
parameter space that mitigates the susceptibility to flux noise without 
sacrificing protection against charge noise and dielectric loss. 
Interestingly, this regime lies beyond the single-band BO description and 
overcomes the coherence limits previously assumed in the 
literature~\cite{messelot2026}.

\section{Hamiltonians}
In this section, we present the full three-mode Hamiltonian of the circuit 
depicted in Fig.~\ref{fig:fig1}a, from which we derive an effective 
single-band BO model, subsequently generalized to include multiple higher-energy bands. Numerical simulations demonstrate that this generalization 
quantitatively captures the low-energy spectrum of the full Hamiltonian.

\subsection{Full three-mode circuit Hamiltonian}
The circuit of Fig.~\ref{fig:fig1}a, biased at offset charge $N_g$ and threaded by external flux $\varphi_\text{ext}$, is described by the following Hamiltonian
\begin{multline}  
{\hat H}_\text{full} = 4E_{C_S} (\hat N-N_g)^2 + 4 E_{C_J} (\hat n_L^2+\hat n_R^2)  \\
- E_J \left(\cos\hat \theta_L + \cos\hat \theta_R \right)+\frac{1}{2}E_L \Big[ (\hat \theta_L - \hat \varphi )^2 + (\hat \theta_R - \hat \varphi + \phiext)^2\Big]
\label{eq:Hfull}
\end{multline}
where $\hat\varphi$ is the phase drop operator across the large capacitor and $\hat N$ its conjugate Cooper-pair number operator. The phase drop operators across the left and right junctions are denoted $\hat\theta_{L/R}$, with $\hat n_{L/R}$ their conjugate Cooper-pair number operators. We assume symmetric interferometer arms throughout, although asymmetric configurations are a straightforward extension. Note that while this gauge is well adapted for the BO analysis that follows, the Hamiltonian is not $2\pi$-periodic in $\hat\varphi$, and therefore the conjugate operator $\hat N$ does not have a discrete spectrum. Consequently, the numerical diagonalization of the circuit is more efficiently carried out in a gauge where the phase $\hat\varphi$ is compact, and its conjugate charge discrete (see Ref.~\cite{Roverch2026} Appendix F.).


\subsection{Single-band BO}
We employ the BO approximation, leveraging the smallness of the capacitive 
ratio $\epsilon = E_{C_S}/E_{C_J} \ll 1$. The $(\hat{N}, \hat{\varphi})$ 
mode plays the role of the heavy nucleus, and $(\hat{n}_{L/R}, 
\hat{\theta}_{L/R})$ those of two light electrons (Fig.~\ref{fig:fig1}b). The BO approximation proceeds 
as follows: one freezes the heavy mode at a classical value $\varphi$ and 
diagonalizes the Hamiltonian of the light modes. The resulting 
$\varphi$-dependent eigenvalues and eigenstates then give rise to an 
effective confining potential for the heavy mode.

The full Hamiltonian may be recast as
\begin{equation}
{\hat H}_\text{full} = 4E_{C_S} (\hat N-N_g)^2 + {\hat H}_\mathrm{light}(\hat\varphi, \phiext)\;.
\label{eq:Hfull_to_Hlight}
\end{equation}
The light-mode Hamiltonian takes the form of two uncoupled 
fluxonium circuits, in which the heavy-mode coordinate ${\varphi}$ acts 
as an effective flux bias, threading the two loops by $\varphi$ and 
$\varphi - \varphi_\text{ext}$ respectively:
\begin{equation}
    {\hat H}_\mathrm{light}(\varphi, \phiext) = \hat H_\mathrm{fluxo}^L (\varphi)+ \hat H_\mathrm{fluxo}^R(\varphi - \phiext) \ ,
\end{equation}
where the fluxonium Hamiltonians are written as
\begin{equation}
    \hat H_\mathrm{fluxo}^{L/R} (\alpha) = 4 E_{CJ} \hat n_{L/R}^2
- E_J \cos\hat \theta_{L/R}+\frac{1}{2}E_L  (\hat \theta_{L/R} - \alpha )^2
\label{eq:Hfluxo}
\end{equation}
for an arbitrary phase bias $\alpha$, with eigenenergies $\left(E_n^{L/R}(\alpha) \right)_{n\in \mathbb{N}}$ and eigenstates $\left(\ket{\psi_n^{L/R}}(\alpha) \right)_{n\in \mathbb{N}}$. For the symmetric circuit considered here $E_n^L = E_n^R$ and $\ket{\psi_n^{L}}=\ket{\psi_n^{R}}$, but we preserve the superscript notation for clarity. Retaining the lowest energy band for the light modes, the resulting single-band BO Hamiltonian reads
\begin{equation}
    \hat{H}_\mathrm{BO,1-band}= 4E_{C_S} (\hat N  - N_g)^2 + U_\mathrm{light}(\hat \varphi) + U_\mathrm{geom}(\hat \varphi)\;,
    \label{H_BO}
\end{equation}
where the usual confining potential arising from the light modes' ground state dispersion is
\begin{equation}
    U_\mathrm{light}(\varphi) = E_0^L(\varphi) + E_0^R(\varphi - \phiext) \ ,
    \label{eq:Ulight}
\end{equation}
and the geometric correction
\begin{align}
    U_\mathrm{geom}( \varphi) = -4E_{C_S} \Big( &\braket{\psi_0^\mathrm{L}}{\partial_\varphi ^2\psi_0^\mathrm{L}}(\varphi) \notag \\
    &+ \braket{\psi_0^\mathrm{R}}{\partial_\varphi ^2\psi_0^\mathrm{R}}(\varphi - \phiext) \Big) \ .
    \label{eq:Ugeom}
\end{align}
Although each $L/R$ contribution to $U_\mathrm{light}$ and $U_\mathrm{geom}$ is individually $2\pi$-periodic in $\varphi$ --- since the underlying energies and eigenstate overlaps are $2\pi$-periodic --- their sum becomes $\pi$-periodic at $\phiext = \pi$ (Fig.~\ref{fig:fig1}d), revealing the Cooper-pair parity of the symmetric interferometer. Interestingly, while the BO approximation has been applied to similar two-Cooper-pair interferometer circuits in other works \cite{Hays2025,Jakobsen2026}, the geometric correction $U_\mathrm{geom}$ was not included. Note that in our parameter regime, its contribution to the potential is significant (Fig.~\ref{fig:fig1}d).
We gain insight on our single-band model at $\phiext=\pi$ by considering the parameter regime of Table~\ref{table1} where $E_J\gtrsim E_{C_J} \gg E_L$. The energy dispersion $E_0^{L/R}(\alpha)$ resembles a set of truncated parabolas of height $\pi^2 E_L /2$, separated from the next band $E_1^{L/R}(\alpha)$ by the so-called BO gap $\Delta$ (Fig.~\ref{fig:fig1}d). Here, $\Delta$ is of the order of the junction phase-slip energy $\varepsilon_\mathrm{ps}$ \cite{Vakhtel2024,Matveev2002}:
\begin{equation}
    \Delta \approx \varepsilon_\mathrm{ps} = \frac{2^{15/4}}{\sqrt{\pi}} E_J^{3/4} E_{C_J}^{1/4} \exp \left( -\sqrt{8\frac{E_J}{E_{C_J}}}\right) \ .
\end{equation}
In the regime where $E_L\gg \varepsilon_\mathrm{ps}$, the peak-to-peak height of the confining potential $U_\mathrm{light}$ is well approximated by $\frac{1}{2}E_L(\pi^2-2(\pi/2)^2)=E_L\pi^2/4$. Decomposing the potential in its Fourier series $U_\mathrm{light}(\varphi)=\sum_n(E_{J2n}\cos(2n\varphi))$ we find
\begin{equation}
    E_{J2n} \underset{\varepsilon_\mathrm{ps} \to 0}{=}\frac{E_L}{n^2}\;.
\end{equation}
Conversely, going to the blochnium regime \cite{Pechenezhskiy2020}, defined as $E_L \ll \varepsilon_\mathrm{ps}$, makes the fluxonium degrees of freedom insensitive to flux offsets, suppressing the confining potentials $U_\mathrm{light}$ and $U_\mathrm{geom}$ (see \ref{sec:blochnium}).

The single-band BO model presented so far discards energy terms of order $\epsilon^2 E_{C_J}^2/\Delta=E_{C_S}^2/\Delta$. In the following, we capture higher-order corrections by devising a BO model including multiple energy bands for the light modes.

\subsection{Multi-band BO}
\label{sec:multiband}
The single-band BO described above projects both light particles in their ground state. This lowest band is therefore labeled $00$. The next step is then to consider either light particle in its first excited state, yielding two bands labeled $01$ and $10$. In this three-band model, the system wavefunction is described by a spinor $\overrightarrow{\ket{\Phi}}=\left(\ket{\Phi_{00}}, \ket{\Phi_{10}}, \ket{\Phi_{01}}\right)$, where each component is a wavefunction of the heavy variable $\varphi$, and the subscript indicates the band index of the light modes. The Hamiltonian is then described by a $3\times 3$ matrix, where each entry is an operator on the heavy variable $\varphi$. Diagonal entries describe a heavy particle of kinetic energy $E_{C_S}$ evolving in the corresponding energy band with geometric corrections (Fig.~\ref{fig:fig1}e). Off-diagonal entries give rise to inter-band couplings. Full expressions of these terms and the generalization to $n$ bands are described in \ref{sec:multiBO}.
In practice, we actually keep $n=k^2$ bands when carrying out numerical computations, so as to keep a tensor structure for the two light modes. In the case of the 3-band model, the added 4th band (11) lies well above the other 3 and has negligible impact on the low energy levels.


\subsection{Comparison of 3- and 1- band models}
\begin{figure}
    \centering
    \includegraphics[scale=1]{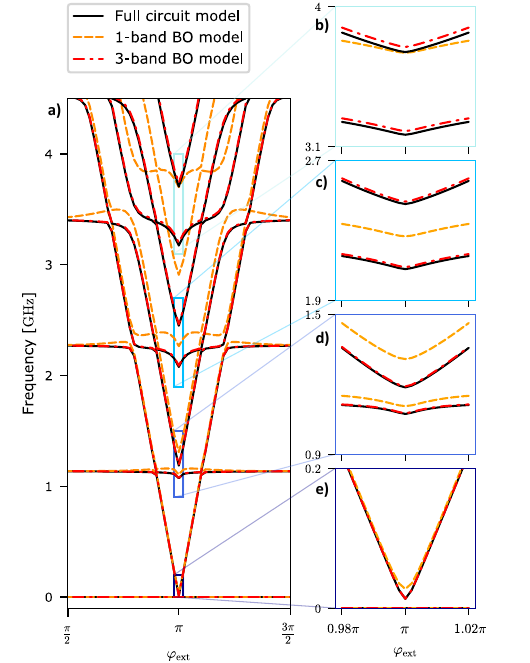}
    \caption{{\bf Benchmark of the Born-Oppenheimer models.} (a) Energy levels (y-axis) versus external flux (x-axis), computed with the full 3-mode lumped circuit model (black), the single-band BO model (orange) and the 3-band BO model (red).
    Panels (b-e) zoom on the first 4 doublets in the vicinity of $\phiext = \pi$. The 3-band model quantitatively captures the full-model energies (3\% error) over the entire band.}
    \label{fig:benchmark_numerics}
\end{figure}

In this section we benchmark our BO models to the full 3-mode circuit Hamiltonian of Eq.~\eqref{eq:Hfull}, with the parameters of Table.~\ref{table1} at $N_g=0$. Transition frequencies from the ground state in the $0-4.5$~GHz band are plotted as a function of $\phiext$ in Fig.~\ref{fig:benchmark_numerics}. Away from $\phiext=\pi$ (e.g. $\phiext=\pi/2, 3\pi/2$ in Fig.~\ref{fig:benchmark_numerics}), we recognize plasmon-like transitions, due to their weak flux dispersion and quasi-harmonic distribution. At $\phiext=\pi$, these plasmons split into doublets of opposite Cooper-pair parity (see insets around the first 4 plasmon transitions in Fig.~\ref{fig:benchmark_numerics}). The single-band BO model of Eq.~\eqref{H_BO} matches the full circuit model for the ground state doublet (inset e), although a significant discrepancy appears exactly at $\phiext=\pi$. The precision degrades as we climb the plasmonic ladder (inset d), and the higher doublet transition appears out of the frame for the second and third plasmon doublets (insets c,b). By contrast, the 3-band BO model of Eq.~\eqref{BO_multiband} is in quantitative agreement with the full circuit model over the full $0-4.5$~GHz, with a maximal relative error of $3\%$. The bandwidth over which the multi-band BO model agrees with the full Hamiltonian can be further extended by including more energy bands (see \ref{sec:multiBO}).

Note that the multi-band BO model is considerably less computationally demanding to diagonalize than the 3-mode circuit Hamiltonian; indeed, for a level truncation per mode $N_\mathrm{max}$, the Hilbert space for the full model is of dimension $N_\mathrm{max}^3$, while with $m$ bands in the BO framework, the final Hilbert space is only of dimension $mN_\mathrm{max}$. Note that other numerical schemes leverage sparsity and the tensor structure of the large Hilbert to efficiently diagonalize multi-mode Hamiltonians.
\begin{table}[h]
\begin{ruledtabular}
\begin{tabular}{ccccc}
 \multicolumn{5}{c}{Lumped-element circuit parameters} \\\cline{1-5}
$E_J \text{[GHz]}$ &$E_{C_J} \text{[GHz]}$& $E_L \text{[GHz]}$ &$E_{C_S} \text{[MHz]}$ \\
\hline
16.83 & $4.82$ & $1.27$ & $72$  \\

\end{tabular}

\end{ruledtabular}
\caption{Circuit parameters of the KITE transmon, corresponding to the measured values in \cite{Roverch2026}.}
\label{table1}
\end{table}

\subsection{Noise operators}
The BO procedure can be formalized by expanding the wavefunction of the multi-mode lumped-element circuit in a basis of states indexed by the continuous heavy-mode variable $\varphi$ together with two discrete quantum numbers labeling the eigenstates of the two light modes (see \ref{sec:multiBO}). This basis is well suited to the problem: including an increasing number of light-mode eigenstates improves the precision in the small parameter $\epsilon$.

In the previous sections, an effective Hamiltonian is obtained by projecting the full circuit Hamiltonian in this basis truncated to the lowest light-modes eigenstate (1-band), and truncated to the three lowest-lying eigenstates (3-band). In this section, we focus on the 1-band model at $\phiext=\pi$ and project all operators that couple to local noise sources.

Dielectric loss in the large capacitor and the junctions couples to $\hat N$ and $\hat n_L, \hat n_R$ respectively while flux noise and inductive loss in the inductors couple to $\hat\phi_{L/R} = \hat\theta_{L/R}-\hat\varphi$. For simplicity, we neglect quasiparticle poisoning, which occurs on timescales much longer than those set by flux noise and dielectric loss \cite{Roverch2026}. For convenience, we introduce $\hat\phi_\Sigma = \frac{1}{2}(\hat\phi_{L}+ \hat\phi_{R})$ and $\hat\phi_\Delta = \frac{1}{2}(\hat\phi_{L}- \hat\phi_{R})$, and $\hat n_\Sigma, \hat n_\Delta$ their respective conjugate variables.  In the following the subscript $\rvert_\mathrm{BO}$ is added to an operator to denote its projection in the single-band model.

The island charge operator can be decomposed using the standard relation \cite{Landau1980, Koch2009}:
\begin{equation}
    \hat N = -i\hat \partial_\varphi + \mathcal{\hat N}^\mathrm{interband}\;,
    \label{Nop}
\end{equation}
where $\mathcal{\hat N}^\mathrm{interband}$ couples between different bands of the fast modes and therefore we simply get 
\begin{equation}
\hat N \rvert_\mathrm{BO} = -i\hat \partial_\varphi\;.
\end{equation}
The Cooper-pair number operators associated with each junction project as
\begin{eqnarray}
    \hat n_{L} \rvert_\mathrm{BO} &=&  \bra{\psi_0^\mathrm{L}} \hat n_L\ket{\psi_0^\mathrm{L}}(\hat \varphi) = 0\;, \\
    \hat n_R \rvert_\mathrm{BO} &=&  \bra{\psi_0^\mathrm{R}} \hat n_R\ket{\psi_0^\mathrm{R}}(\hat \varphi-\pi) = 0\;. 
\end{eqnarray}
The symmetric phase drop across the inductors projects to
\begin{equation}
    \hat \phi_\Sigma \rvert_\mathrm{BO} = \frac{1}{2} \left( \bra{\psi_0^\mathrm{L}} \hat \phi_L  \ket{\psi_0^\mathrm{L}}(\hat \varphi)  + \bra{\psi_0^\mathrm{R}} \hat \phi_R \ket{\psi_0^\mathrm{R}}(\hat \varphi - \pi)\right)
\end{equation}
which is a $\pi$-periodic function of $\varphi$, and hence corresponds to two-Cooper-pair tunneling. And finally the anti-symmetric phase drop across the inductors projects to
\begin{equation}
    \hat \phi_\Delta \rvert_\mathrm{BO} = \frac{1}{2} \left( \bra{\psi_0^\mathrm{L}} \hat \phi_L\ket{\psi_0^\mathrm{L}}(\hat \varphi)  - \bra{\psi_0^\mathrm{R}} \hat \phi_R\ket{\psi_0^\mathrm{R}}(\hat \varphi - \pi)\right)\;,
\end{equation}
which is $2\pi$-periodic function of $\varphi$, and therefore corresponds to single Cooper-pair tunneling, breaking the Cooper-pair-parity symmetry.

In the next section, we analyze in greater detail which symmetries these noise operators break or preserve, and how this impacts the lifetime of the qubit encoded in the lowest-lying doublet of states.

\section{Symmetries}

\subsection{Symmetry operators}
\subsubsection{Single-mode}
The ideal $\cos 2\varphi$ Hamiltonian \eqref{eq:Hcos2phi} at zero offset charge features two distinct reflection symmetries
\begin{equation}
     \mathcal{S}_1 :{\varphi} \longmapsto -{\varphi}, \quad \mathcal{P}_1 : {\varphi} \longmapsto \pi - {\varphi} \ .
     \label{symmetries1}
\end{equation}
Since the eigenspectrum of fluxonium is symmetric around $\phiext= 0,\pi$, $U_\mathrm{light}$ and $ U_\mathrm{geom}$ have reflection symmetry around $\varphi=0,\pi/2$ (see Fig.~\ref{fig:fig1}d), and therefore $\mathcal{S}_1$ and $\mathcal{P}_1$ symmetries also hold for the Born-Oppenheimer reduced Hamiltonian $\hat{H}_\mathrm{BO,1-band}$ \eqref{H_BO} at 0 offset charge. A question then arises: how are these symmetries affected when we introduce imperfections? Let's first introduce $L/R$ asymmetry in the KITE, while maintaining $\phiext=\pi$. The $L/R$ contributions in Eqs.~\eqref{eq:Ulight},\eqref{eq:Ugeom} are unequal, although each term preserves its reflection symmetry. Therefore, $\mathcal{S}_1$ symmetry is preserved, but $\mathcal{P}_1$ symmetry is broken. Conversely, restoring a perfectly symmetric KITE but assuming $\phiext$ deviates from $\pi$ breaks $\mathcal{S}_1$ symmetry, but reflection symmetry around $\phiext$: $\mathcal{P}_1(\phiext) = {\varphi} \longmapsto \phiext - \varphi$ is preserved. 


Let us now consider the Cooper-pair-pairing property, that corresponds to a $\pi$-periodicity symmetry $\mathcal{T}_1: {\varphi} \longmapsto {\varphi + \pi}$. Interestingly, $\mathcal{T}_1 = \mathcal{S}_1 \mathcal{P}_1 = \mathcal{P}_1 \mathcal{S}_1$, and therefore $\mathcal{T}_1$ symmetry arises for a symmetric KITE at $\phiext=\pi$. The eigenvalue equation of a particle evolving in a $\pi-$periodic potential ($\mathcal{T}_1$ symmetry) is known as the Hill equation \cite{Magnus2013}, for which general properties are known. Namely, its eigenvalues may be sorted as follows:
\begin{equation}
    E_{0+} < E_{0-} < E_{1-} < E_{1+} < \dots \ ,
\end{equation}
where the $\pm$ index corresponds to the parity of the $\mathcal{T}_1$ operator, and the quantum number $n=0,1,\dots$ is assigned to each even/odd doublet. More formally, denoting $\ket{n\pm}$ the corresponding eigenstates, we have
\begin{equation*}
    \mathcal{T}_1 \ket{n\pm} = \pm \ket{n \pm}\;.
\end{equation*}

In addition, the $\mathcal{P}_1$ and $\mathcal{S}_1$ operators are involutive, so the Hamiltonian eigenstates must also fall into even or odd parity sectors for those two symmetries. For the lowest set of eigenvalues and the parameter regime explored in this paper (Table~\ref{table1}), the following equalities hold
\begin{align}
    \mathcal{P}_1 \ket{n \pm} &= (-1)^n \ket{n \pm} \\
    \mathcal{S}_1 \ket{n \pm} &= \pm(-1)^n \ket{n \pm} \ .
\end{align}

\subsubsection{Three-mode}
Similar symmetries can be found in the full Hamiltonian \eqref{eq:Hfull} at zero offset charge : 

\begin{equation}
   \mathcal{S}_3: \left\{ \begin{aligned} {\varphi} &\longmapsto-{\varphi} \\
        {\theta_L} &\longmapsto -{\theta_L} \\
        {\theta_R} &\longmapsto -2\pi -{\theta_R} \\
        \end{aligned} \right. , \quad 
        \mathcal{P}_3(\phiext):\left\{ \begin{aligned} {\varphi} &\longmapsto \phiext-{\varphi} \\
        {\theta_L} &\longmapsto -{\theta_R} \\
        {\theta_R} &\longmapsto  -{\theta_L} \\
        \end{aligned} \right.
\end{equation}
Once again we have $[\mathcal{\hat H}, \mathcal{S}_3] = 0$ regardless of asymmetry at $\phiext=\pi$, and $[\mathcal{\hat H}, \mathcal{P}_3] = 0 $ for all $ \phiext$ for a symmetric circuit. Note that Cooper-pair-parity symmetry is not explicit in this gauge (chosen for the BO approximation), and is instead apparent in the gauge best suited for numerical diagonalization (see Ref.~\cite{Roverch2026} Appendix F.).  

We now proceed to the eigenstate labeling of the full three-mode Hamiltonian. Note that as circuit parameters are varied, the assignment of eigenstates into parity sectors remains constant as long as eigenvalues do not cross. Therefore, we may reduce $E_{C_S}$ keeping all other parameters fixed, thereby reducing the capacitive ratio $\epsilon=E_{C_S}/E_{C_J}$, and hence the spectrum of the full three-mode Hamiltonian converges to that of the single-mode BO model. We can then allocate the labeling derived for the single-mode BO model. In our parameter regime, and for the lowest eigenstates considered here, no eigenvalue crossings are observed when restoring the nominal value of $E_{C_S}$. We therefore preserve
\begin{align}
    \mathcal{P}_3 \ket{n \pm} &= (-1)^n \ket{n \pm} \\
    \mathcal{S}_3 \ket{n \pm} &= \pm(-1)^n \ket{n \pm} \ .
\end{align}

\begin{figure}
    \centering
    \includegraphics[scale=1]{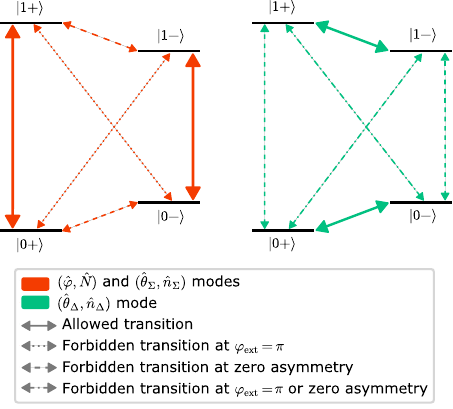}
    \caption{ {\bf Selection rules.} Energy level diagram for the 4 lowest eigenstates (not to scale). Double arrows indicate couplings between levels through circuit operators and the linestyle encodes the selection rules derived in Eqs.~\ref{eq:P_slr},\ref{eq:S_slr}. The ground state doublet protection is preserved by ($\hat N, \hat \varphi, \hat n_{\Sigma}, \hat \theta_{\Sigma}$) in red, while it is broken by ($\hat n_{\Delta}, \hat \theta_{\Delta}$) in green.}
    \label{fig:fig4}
\end{figure}

\subsection{Selection rules}
The symmetries of the circuit Hamiltonian provide invaluable information on selection rules for transitions between eigenstates through the action of any operator. Namely, let us consider two eigenstates $\ket{v_1}, \ket{v_2}$, an operator $\hat O$ and a symmetry operator $\mathcal{S}$. The matrix element vanishes by symmetry: $\bra{v_1}\hat O\ket{v_2}=0$ if (1) $\ket{v_1}, \ket{v_2}$ lie in the same parity sector for $\mathcal{S}$ and the operators anti-commute: $\{\mathcal{S},\hat O\}=0$ or (2) $\ket{v_1}, \ket{v_2}$ lie in opposite parity sectors for $\mathcal{S}$ and the operators commute: $[\mathcal{S},\hat O]=0$. A slight generalization of these statements is that the commutator or anti-commutator need not be exactly 0; it suffices that they are diagonal in the eigenbasis.

To that end, we compute the following commutation and anti-commutation relations:
\begin{equation}
 \{ \mathcal{S}_3, {\hat N} \} = \{ \mathcal{S}_3, {\hat n_{\Sigma}} \} = \{ \mathcal{S}_3, {\hat n_{\Delta}} \} = \{ \mathcal{S}_3, {\hat \varphi} \} = 0
 \end{equation}
\begin{equation}
 -\{ \mathcal{S}_3, {\hat \theta_{\Sigma}} \} = \{ \mathcal{S}_3, {\hat \theta_{\Delta}} \} = \pi \mathcal{S}_3
 \end{equation}
and
\begin{equation}
 \{ \mathcal{P}_3, {\hat N} \} = \{ \mathcal{P}_3, {\hat n_{\Sigma}} \} = \{ \mathcal{P}_3,{\hat \theta_{\Sigma}} \} = 0
 \end{equation}
 \begin{equation}
     \{ \mathcal{P}_3, {\hat \varphi} \} = \phiext \mathcal{P}_3
 \end{equation}
 \begin{equation}
     [\mathcal{P}_3, {\hat n_{\Delta}} ] = [\mathcal{P}_3, {\hat \theta_{\Delta}} ] = 0\;.
 \end{equation}
 This leads to the following selection rules: when $\phiext=\pi$, $\mathcal{S}_3-$symmetry yields:
 \begin{equation}
      \begin{aligned} 
  n-m \in 2\mathbb{Z} \setminus \{0\} \Rightarrow \left\{ \begin{aligned} 
  \bra{m\pm} {\hat N} \ket{n \pm}=\bra{m,\pm} {\hat {\varphi}} \ket{n \pm}  &= 0 \\
  \bra{m\pm} {n_{\Sigma}} \ket{n\pm}=\bra{m\pm} {\theta_{\Sigma}} \ket{n \pm}  &= 0 \\\bra{m\pm} {\hat n_{\Delta}} \ket{n \pm}=\bra{m\pm} {\hat \theta_{\Delta}} \ket{n \pm}  &= 0 \\
  \end{aligned} \right.\\
  n-m \notin 2\mathbb{Z} \Rightarrow \left\{ \begin{aligned} 
  \bra{m\pm} {\hat N} \ket{n \mp}=\bra{m\pm} {\hat {\varphi}} \ket{n \mp}  &= 0 \\
  \bra{m\pm} {\hat n_{\Sigma}} \ket{n \mp}=\bra{m\pm} {\hat \theta_{\Sigma}} \ket{n \mp}  &= 0 \\\bra{m\pm} {\hat n_{\Delta}} \ket{n \mp}=\bra{m\pm} {\hat \theta_{\Delta}} \ket{n \mp}  &= 0 \\
  \end{aligned} \right.
\end{aligned} \label{eq:S_slr}
\end{equation}
while for a symmetric KITE, $\mathcal{P}_3-$symmetry yields:
\begin{equation}
      \begin{aligned} 
  n-m \in 2\mathbb{Z} \setminus \{0\} \Rightarrow \left\{ \begin{aligned} 
  \bra{m\pm} {\hat N} \ket{n \pm}=\bra{m\pm} {\hat {\varphi}} \ket{n \pm}  &= 0 \\
  \bra{m,\pm} {\hat n_{\Sigma}} \ket{n \pm}=\bra{m\pm} {\hat \theta_{\Sigma}} \ket{n \pm}  &= 0 \\
  \end{aligned} \right.\\
  n-m \in 2\mathbb{Z} \Rightarrow \left\{ \begin{aligned} 
  \bra{m\pm} {\hat N} \ket{n \mp}=\bra{m\pm} {\hat {\varphi}} \ket{n \mp}  &= 0 \\
  \bra{m\pm} {\hat n_{\Sigma}} \ket{n \mp}=\bra{m\pm} {\hat \theta_{\Sigma}} \ket{n \mp}  &= 0 \\
  \end{aligned} \right.\\
  n-m \notin 2\mathbb{Z} \Rightarrow \left\{ \begin{aligned} 
  \bra{m\pm} {\hat n_{\Delta}} \ket{n \pm}=\bra{m\pm} {\hat \theta_{\Delta}} \ket{n \pm}  &= 0 \\
   \bra{m\pm} {\hat n_{\Delta}} \ket{n \mp}=\bra{m\pm} {\hat \theta_{\Delta}} \ket{n \mp}  &= 0 \\
  \end{aligned} \right.
\end{aligned} \label{eq:P_slr}
\end{equation}

These selection rules are summarized for the first 4 eigenstates in Fig.\ref{fig:fig4}. Note that the symmetries of the one-mode model \eqref{symmetries1} lead to the same selection rules for operators $\hat N \rvert_\mathrm{BO}, \hat \theta_\Sigma \rvert_\mathrm{BO}, \hat \theta_\Delta \rvert_\mathrm{BO}$.

One may conclude from this analysis that while thermally activated leakage out of the $\{\ket{0+},\ket{0-}\}$ doublet towards higher states is exponentially suppressed by the spectral gap, transitions within the doublet are (i) protected against $\{ \hat N, \hat \varphi, \hat n_{\Sigma}, \hat \theta_{\Sigma} \}$ for a symmetric KITE and regardless of deviations from $\phiext=\pi$, (ii) unprotected against the $\{\hat n_\Delta,\hat\theta_\Delta\}$ mode that is the Achilles heel of this qubit.

\section{Coherence}
The previous section addressed selection rules governing $T_1$ errors; here we turn to $T_2$ errors, i.e., dephasing caused by the sensitivity of the qubit frequency $\omega_q$ to noisy external parameters. Two sources dominate in our circuit: (i) fluctuations of the island's offset charge, and (ii) fluctuations of the flux threading the KITE loop. Our qubit operates at the double sweet spot $N_g=0$, $\phiext=\pi$, where $\omega_q$ is first-order insensitive to both. The residual dephasing rate --- from second-order fluctuations or small miscalibrations --- then scales with the curvature of the qubit frequency with respect to the corresponding noisy parameter \cite{Ithier2005}. Hereafter, $\bar\omega_q$ denotes the qubit frequency evaluated at the double sweet spot, $\bar\omega_q \equiv \omega_q(N_g=0,\phiext=\pi)$.

In the transmon regime, the qubit frequency versus offset charge $N_g$ is well approximated by $\omega_q(N_g)\approx\bar\omega_q\abs{\cos(\pi N_g)}$, and so the curvature is entirely determined by the qubit frequency:
\begin{equation}
\frac{\partial^2\omega_q}{\partial N_g^2}\Big\rvert_{N_g=0}=-\pi^2\bar\omega_q\;.    
\end{equation}


We place ourselves in the regime where $E_L$ dominates the phase-slip energy $E_L>\varepsilon_{ps}$ (see \ref{sec:blochnium} for the opposite regime). Away from the sweetspot, the doublet states localize in wells located at $(\varphi, \theta_L, \theta_R)=(\frac{\pi}{2}, 0,0)$ or $(-\frac{\pi}{2},0,-2\pi)$ (Fig.~\ref{fig:blochnium}a). First-order perturbation theory then yields the slope of the qubit frequency versus external flux:
\begin{align*}
    \frac{d (E_{0+}- E_{0-})}{d \phiext} &= \bra{0+} \frac{\partial \hat H_\mathrm{full}}{\partial \phiext} \ket{0+} - \bra{0-} \frac{\partial \hat H_\mathrm{full}}{\partial \phiext} \ket{0-} \\
    &\approx \pm E_L \times \pi \ .
\end{align*}
The qubit frequency versus external flux $\delta\phiext=\phiext-\pi$ is then well approximated by 
$\omega_q(\delta\phiext)\approx\frac{1}{\hbar}\sqrt{(\pi E_L \delta\phiext)^2+(\hbar\bar\omega_q)^2}$, and hence the curvature reads 
\begin{equation}
\frac{\partial^2\omega_q}{\partial \phiext^2}\Big\rvert_{\phiext=\pi}=\pi^2 (E_L/\hbar)^2/\bar\omega_q\;.    
\end{equation}

The procedure of optimizing circuit parameters to maximize dephasing times is constrained: the plasmon frequency $\omega_\mathrm{plasmon}$, which sets the gap to the rest of the spectrum, must satisfy $\hbar\omega_\mathrm{plasmon} > k_B T$, or thermal transitions out of the qubit manifold will proliferate. This frequency may be computed by considering the excitations of the slow $\varphi$ mode within a single well: if the two light modes $\theta_L$, $\theta_R$ are localized in their respective potential minima (this approximation is valid as long as $E_J$ is the dominant energy scale), then the $\varphi$ mode behaves like a harmonic oscillator of charging energy $E_{C_S}$ and inductive energy $2E_L$, leading to a plasmon frequency 
\begin{equation}
\omega_\mathrm{plasmon}=\sqrt{16 E_L E_{C_S}}/\hbar\;.
\end{equation}

\subsection{Coherence limitations in the single-mode regime}
In the parameter regime where the single-mode BO approximation is valid, that is $E_{C_S}\ll E_{C_J}, \Delta$, we have $E_{J2}\approx E_L$ and hence \cite{Smith2020}
\begin{equation}
\bar\omega_q\big\rvert_\mathrm{1-mode} \approx  16 \sqrt{\frac{2}{\pi}} \frac{E_L}{\hbar} \left(\frac{2 E_{C_S}}{E_L}\right)^{1/4}e^{-\sqrt{2 E_L/E_{C_S}}}\;.
\end{equation}
In this regime, the optimization procedure described above rests solely on the two parameters $E_L$ and $E_{C_S}$. One may proceed as follows: (i) set $\bar\omega_q$ in the MHz range so that charge noise dephasing is in the millisecond range. This mostly constrains the ratio $E_L/E_{C_S}$. (ii) Minimizing flux noise dephasing without compromising on charge noise demands reducing $E_L$ while also reducing $E_{C_S}$ in order to keep $\bar\omega_q$ fixed. A sub-100~MHz $E_L$ pushes flux-noise dephasing towards the millisecond range. (iii) This in turn will cause the collapse of the plasmon frequency to the 100~MHz range, causing significant thermal activated processes out of the qubit manifold. The unsatisfiability of these three criteria (low charge-noise and flux-noise dephasing, while keeping a significant plasmon frequency) has been predicted in all interference-based $\cos 2\varphi-$transmon implementations \cite{messelot2026}, and referred to as the flux/charge tradeoff.

\subsection{Beyond the single-mode regime: escaping the flux/charge tradeoff}
\begin{figure}
    \centering
    \includegraphics[scale=1]{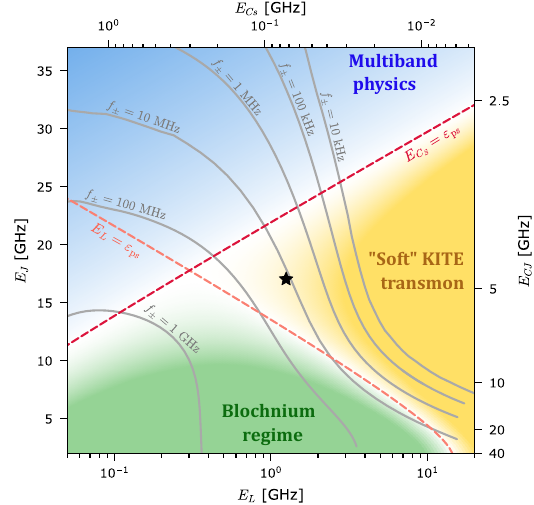}
    \caption{{\bf Phase diagram.} Operation regimes of our circuit (color) and iso-frequencies of the ground state doublet (gray solid lines) are represented in the $(E_L, E_J)$ plane. We keep fixed the products $E_{C_S} \times E_L$ and $E_{C_J} \times E_J$, that control the plasmon and plasma frequencies. The three regimes of interest are separated by boundaries (dashed lines) of equations $E_L = \varepsilon_\mathrm{ps}$ and $E_{C_S} = \varepsilon_\mathrm{ps}$.
    The black star denotes the parameter set of \cref{table1}, corresponding to the experimental implementation of Ref.~\cite{Roverch2026}.
    }
    \label{fig:fig5}
\end{figure}
In this section, we evade the flux/charge tradeoff by moving towards a parameter regime where the single-mode BO approximation breaks down. While we preserve the hierarchy of masses $E_{C_S}\ll E_{C_J}$, we close the BO gap $E_{C_S} \gtrsim \Delta$. In this regime, the system is described by multiple energy bands, as shown in Sec.~\ref{sec:multiband}. We find that as the gap $\Delta$ closes, the higher bands contribute significantly to the confinement of the lowest qubit states, which is no longer governed only by $E_L$. As a result, $\Delta$ acts as an extra knob to control $\bar\omega_q$ and manage charge-noise dephasing. On the other hand, the flux sensitivity is still governed by $E_L$. An optimizing procedure that keeps $\omega_\mathrm{plasmon}$ fixed becomes possible (in Fig.~\ref{fig:fig5} we further fix the junction plasma frequency $\omega_\mathrm{plasma}=\sqrt{8 E_J E_{C_J}}/\hbar$): (i) set a value of $\bar\omega_q$ to define a threshold for charge-noise dephasing. This constrains us to a 1D curve in the $(E_J/E_{C_J}, E_L/E_{C_S})$ plane (see Fig.~\ref{fig:fig5}). (ii) Follow the iso-frequency curve to reduce $E_L$ until flux-noise dephasing is sufficiently suppressed.

In summary, by tuning the energy gap $\Delta$, it appears possible to design a qubit with arbitrarily low charge-noise and flux-noise dephasing, while keeping the plasmon and plasma frequencies fixed. Moreover, reducing $E_L$ also reduces the flux-noise-induced transition rates. Throughout, the symmetry-induced selection rules that protect from dielectric loss remain valid.

\section{Conclusion}
In summary, we have derived a single-band model -- and its multi-band generalization -- from the full three-mode model of a $\cos 2\varphi$ transmon circuit based on the Born-Oppenheimer approximation. These reduced models reveal the mechanism of Cooper-pair pairing of the full circuit, and reduce the complexity of numerical simulations. Starting from the single-band model, and generalizing to the full circuit model, we find symmetry operators that result in selection rules for all circuit dipole operators. This analysis demonstrates our qubit's protection from dielectric loss and its vulnerability to flux noise, a manifestation of the charge/flux tradeoff known in $\cos 2\varphi$ qubits. Remarkably, we then evade this tradeoff by moving towards circuit parameters (namely lowering $E_L$ and closing the BO gap $\Delta$) that strongly involve the higher energy bands in the qubit states dynamics. 

In hindsight, the multi-mode nature of our circuit has posed challenges such as introducing additional decoherence channels, but may also be leveraged as a resource to shape qubit eigenstates into a protected regime.

Looking ahead, multiple paths are to be explored. Moving in the direction of high coherence will most likely eventually lead to a coherence plateau due to an unforeseen decoherence mechanism. We anticipate that exploring this type of protected qubit is a rich path to deepen our understanding of loss mechanisms in superconducting circuits. Finally, going deeper in the protected regime will require addressing the challenges of control and readout, an ongoing conundrum for protected qubits.

\bibliographystyle{apsrev4-1}
\bibliography{biblio}

@article{Koch2007,
  title = {Charge-insensitive qubit design derived from the Cooper pair box},
  author = {Koch, Jens and Yu, Terri M. and Gambetta, Jay and Houck, A. A. and Schuster, D. I. and Majer, J. and Blais, Alexandre and Devoret, M. H. and Girvin, S. M. and Schoelkopf, R. J.},
  journal = {Phys. Rev. A},
  volume = {76},
  issue = {4},
  pages = {042319},
  numpages = {19},
  year = {2007},
  month = {Oct},
  publisher = {American Physical Society},
  doi = {10.1103/PhysRevA.76.042319},
  url = {https://link.aps.org/doi/10.1103/PhysRevA.76.042319}
}

@article{GyenisReview,
  title = {Moving beyond the Transmon: Noise-Protected Superconducting Quantum Circuits},
  author = {Gyenis, Andr\'as and Di Paolo, Agustin and Koch, Jens and Blais, Alexandre and Houck, Andrew A. and Schuster, David I.},
  journal = {PRX Quantum},
  volume = {2},
  issue = {3},
  pages = {030101},
  numpages = {15},
  year = {2021},
  month = {Sep},
  publisher = {American Physical Society},
  doi = {10.1103/PRXQuantum.2.030101},
  url = {https://link.aps.org/doi/10.1103/PRXQuantum.2.030101}
}

@article{Bell2014,
  title = {Protected Josephson Rhombus Chains},
  author = {Bell, Matthew T. and Paramanandam, Joshua and Ioffe, Lev B. and Gershenson, Michael E.},
  journal = {Phys. Rev. Lett.},
  volume = {112},
  issue = {16},
  pages = {167001},
  numpages = {5},
  year = {2014},
  month = {Apr},
  publisher = {American Physical Society},
  doi = {10.1103/PhysRevLett.112.167001},
  url = {https://link.aps.org/doi/10.1103/PhysRevLett.112.167001}
}

@misc{Nguyen2025,
      title={The superconducting grid-states qubit}, 
      author={Long B. Nguyen and Hyunseong Kim and Dat T. Le and Thomas Ersevim and Sai P. Chitta and Trevor Chistolini and Christian Jünger and W. Clarke Smith and T. M. Stace and Jens Koch and David I. Santiago and Irfan Siddiqi},
      year={2025},
      eprint={2509.14656},
      archivePrefix={arXiv},
      primaryClass={quant-ph},
      url={https://arxiv.org/abs/2509.14656}, 
}

@article{Kalashnikov2020,
  title = {Bifluxon: Fluxon-Parity-Protected Superconducting Qubit},
  author = {Kalashnikov, Konstantin and Hsieh, Wen Ting and Zhang, Wenyuan and Lu, Wen-Sen and Kamenov, Plamen and Di Paolo, Agustin and Blais, Alexandre and Gershenson, Michael E. and Bell, Matthew},
  journal = {PRX Quantum},
  volume = {1},
  issue = {1},
  pages = {010307},
  numpages = {15},
  year = {2020},
  month = {Sep},
  publisher = {American Physical Society},
  doi = {10.1103/PRXQuantum.1.010307},
  url = {https://link.aps.org/doi/10.1103/PRXQuantum.1.010307}
}

@article{Gyenis0pi,
  title = {Experimental Realization of a Protected Superconducting Circuit Derived from the $0$--$\ensuremath{\pi}$ Qubit},
  author = {Gyenis, Andr\'as and Mundada, Pranav S. and Di Paolo, Agustin and Hazard, Thomas M. and You, Xinyuan and Schuster, David I. and Koch, Jens and Blais, Alexandre and Houck, Andrew A.},
  journal = {PRX Quantum},
  volume = {2},
  issue = {1},
  pages = {010339},
  numpages = {18},
  year = {2021},
  month = {Mar},
  publisher = {American Physical Society},
  doi = {10.1103/PRXQuantum.2.010339},
  url = {https://link.aps.org/doi/10.1103/PRXQuantum.2.010339}
}

@article{Hays2025,
  title = {Nondegenerate Noise-Resilient Superconducting Qubit},
  author = {Hays, Max and Kim, Junghyun and Oliver, William D.},
  journal = {PRX Quantum},
  volume = {6},
  issue = {4},
  pages = {040321},
  numpages = {19},
  year = {2025},
  month = {Oct},
  publisher = {American Physical Society},
  doi = {10.1103/dd96-gcb6},
  url = {https://link.aps.org/doi/10.1103/dd96-gcb6}
}

@article{Smith2025,
  title={Spectral signature of high-order photon processes enhanced by Cooper-pair pairing},
  author={Smith, WC and Borgognoni, A and Villiers, M and Roverc’h, E and Palomo, J and Delbecq, MR and Kontos, T and Campagne-Ibarcq, P and Dou{\c{c}}ot, B and Leghtas, Z},
  journal={Nature Communications},
  volume={16},
  number={1},
  pages={8359},
  year={2025},
  publisher={Nature Publishing Group UK London}
}

@article{Smith2020,
  title={Superconducting circuit protected by two-Cooper-pair tunneling},
  author={Smith, WC and Kou, A and Xiao, X and Vool, U and Devoret, MH},
  journal={npj Quantum Information},
  volume={6},
  number={1},
  pages={8},
  year={2020},
  publisher={Nature Publishing Group UK London}
}

@article{Smith2022,
  title={Magnifying quantum phase fluctuations with Cooper-pair pairing},
  author={Smith, William C and Villiers, Marius and Marquet, Antoine and Palomo, Jose and Delbecq, MR and Kontos, Takis and Campagne-Ibarcq, Philippe and Dou{\c{c}}ot, Beno{\^\i}t and Leghtas, Zaki},
  journal={Physical Review X},
  volume={12},
  number={2},
  pages={021002},
  year={2022},
  publisher={APS}
}

@article{GKP2001,
  title = {Encoding a qubit in an oscillator},
  author = {Gottesman, Daniel and Kitaev, Alexei and Preskill, John},
  journal = {Phys. Rev. A},
  volume = {64},
  issue = {1},
  pages = {012310},
  numpages = {21},
  year = {2001},
  month = {Jun},
  publisher = {American Physical Society},
  doi = {10.1103/PhysRevA.64.012310},
  url = {https://link.aps.org/doi/10.1103/PhysRevA.64.012310}
}

@article{Brosco2024flowermon,
  title={Superconducting qubit based on twisted cuprate van der Waals heterostructures},
  author={Brosco, Valentina and Serpico, Giuseppe and Vinokur, Valerii and Poccia, Nicola and Vool, Uri},
  journal={Physical Review Letters},
  volume={132},
  number={1},
  pages={017003},
  year={2024},
  publisher={APS}
}

@article{Messelot2024,
  title={Direct measurement of a sin (2 $\varphi$) current phase relation in a graphene superconducting quantum interference device},
  author={Messelot, Simon and Aparicio, Nicolas and De Seze, Elie and Eyraud, Eric and Coraux, Johann and Watanabe, Kenji and Taniguchi, Takashi and Renard, Julien},
  journal={Physical Review Letters},
  volume={133},
  number={10},
  pages={106001},
  year={2024},
  publisher={APS}
}

@article{Larsen2020,
  title={Parity-protected superconductor-semiconductor qubit},
  author={Larsen, Thorvald W and Gershenson, Michael E and Casparis, Lucas and Kringh{\o}j, Anders and Pearson, Natalie J and McNeil, Robert PG and Kuemmeth, Ferdinand and Krogstrup, Peter and Petersson, Karl D and Marcus, Charles M},
  journal={Physical review letters},
  volume={125},
  number={5},
  pages={056801},
  year={2020},
  publisher={APS}
}

@misc{messelot2026,
      title={Coherence Limits in Interference-Based cos(2$\varphi$) Qubits}, 
      author={S. Messelot and A. Leblanc and J. -S. Tettekpoe and F. Lefloch and Q. Ficheux and J. Renard and E. Dumur},
      year={2026},
      eprint={2601.10209},
      archivePrefix={arXiv},
      primaryClass={quant-ph},
      url={https://arxiv.org/abs/2601.10209}, 
}

@misc{Roverch2026,
      title={Experimental realization of a $\cos(2\varphi)$ transmon qubit}, 
      author={Erwan Roverc'h and Alvise Borgognoni and Marius Villiers and Kyrylo Gerashchenko and W. Clarke Smith and Christopher Wilson and Benoit Douçot and Alexandru Petrescu and Philippe Campagne-Ibarcq and Zaki Leghtas},
      year={2026},
      eprint={2603.13114},
      archivePrefix={arXiv},
      primaryClass={quant-ph},
      url={https://arxiv.org/abs/2603.13114}, 
}

@article{Gladchenko2008,
  title = {Superconducting nanocircuits for topologically protected qubits},
  volume = {5},
  ISSN = {1745-2481},
  url = {http://dx.doi.org/10.1038/nphys1151},
  DOI = {10.1038/nphys1151},
  number = {1},
  journal = {Nature Physics},
  publisher = {Springer Science and Business Media LLC},
  author = {Gladchenko,  Sergey and Olaya,  David and Dupont-Ferrier,  Eva and Dou\c{c}ot,  Benoit and Ioffe,  Lev B. and Gershenson,  Michael E.},
  year = {2008},
  month = nov,
  pages = {48–53}
}

@article{Dodge2023,
  title = {Hardware Implementation of Quantum Stabilizers in Superconducting Circuits},
  author = {Dodge, K. and Liu, Y. and Klots, A. R. and Cole, B. and Shearrow, A. and Senatore, M. and Zhu, S. and Ioffe, L. B. and McDermott, R. and Plourde, B. L. T.},
  journal = {Phys. Rev. Lett.},
  volume = {131},
  issue = {15},
  pages = {150602},
  numpages = {6},
  year = {2023},
  month = {Oct},
  publisher = {American Physical Society},
  doi = {10.1103/PhysRevLett.131.150602},
  url = {https://link.aps.org/doi/10.1103/PhysRevLett.131.150602}
}

@article{Ioffe2002,
  title = {Possible realization of an ideal quantum computer in Josephson junction array},
  author = {Ioffe, L. B. and Feigel'man, M. V.},
  journal = {Phys. Rev. B},
  volume = {66},
  issue = {22},
  pages = {224503},
  numpages = {8},
  year = {2002},
  month = {Dec},
  publisher = {American Physical Society},
  doi = {10.1103/PhysRevB.66.224503},
  url = {https://link.aps.org/doi/10.1103/PhysRevB.66.224503}
}

@article{Doucot2002,
  title = {Pairing of Cooper Pairs in a Fully Frustrated Josephson-Junction Chain},
  author = {Dou\ifmmode \mbox{\c{c}}\else \c{c}\fi{}ot, Benoit and Vidal, Julien},
  journal = {Phys. Rev. Lett.},
  volume = {88},
  issue = {22},
  pages = {227005},
  numpages = {4},
  year = {2002},
  month = {May},
  publisher = {American Physical Society},
  doi = {10.1103/PhysRevLett.88.227005},
  url = {https://link.aps.org/doi/10.1103/PhysRevLett.88.227005}
}

@article{Groszkowski2018,
  title={Coherence properties of the 0-$\pi$ qubit},
  author={Groszkowski, Peter and Paolo, A Di and Grimsmo, AL and Blais, A and Schuster, DI and Koch, Jens},
  journal={New Journal of Physics},
  volume={20},
  number={4},
  pages={043053},
  year={2018},
  publisher={IOP Publishing}
}

@article{Ardati2024,
  title = {Using Bifluxon Tunneling to Protect the Fluxonium Qubit},
  author = {Ardati, Wa\"el and L\'eger, S\'ebastien and Kumar, Shelender and Suresh, Vishnu Narayanan and Nicolas, Dorian and Mori, Cyril and D'Esposito, Francesca and Vakhtel, Tereza and Buisson, Olivier and Ficheux, Quentin and Roch, Nicolas},
  journal = {Phys. Rev. X},
  volume = {14},
  issue = {4},
  pages = {041014},
  numpages = {15},
  year = {2024},
  month = {Oct},
  publisher = {American Physical Society},
  doi = {10.1103/PhysRevX.14.041014},
  url = {https://link.aps.org/doi/10.1103/PhysRevX.14.041014}
}

@article{Mencia2024,
  title = {Integer Fluxonium Qubit},
  author = {Mencia, Raymond A. and Lin, Wei-Ju and Cho, Hyunheung and Vavilov, Maxim G. and Manucharyan, Vladimir E.},
  journal = {PRX Quantum},
  volume = {5},
  issue = {4},
  pages = {040318},
  numpages = {12},
  year = {2024},
  month = {Nov},
  publisher = {American Physical Society},
  doi = {10.1103/PRXQuantum.5.040318},
  url = {https://link.aps.org/doi/10.1103/PRXQuantum.5.040318}
}

@misc{Zhurbina2026,
      title={Coherence limitations of a Fourier-engineered $\cos(2\varphi)$ transmon qubit}, 
      author={Nataliia K. Zhurbina and Siddharth Singh and Lukas J. Splitthoff and Eugene Y. Huang and Figen Yilmaz and A. Mert Bozkurt and Christian Kraglund Andersen},
      year={2026},
      eprint={2605.06372},
      archivePrefix={arXiv},
      primaryClass={quant-ph},
      url={https://arxiv.org/abs/2605.06372}, 
}

@misc{Sanchez2026,
      title={Revisiting the multi-mode rhombus circuit as a biased-noise qubit}, 
      author={Pablo Aramburu Sanchez and Trevyn F. Q. Larson and Anthony P. McFadden and Constantin Schrade and Joshua Combes and András Gyenis},
      year={2026},
      eprint={2605.06430},
      archivePrefix={arXiv},
      primaryClass={quant-ph},
      url={https://arxiv.org/abs/2605.06430}, 
}

@misc{Jakobsen2026,
      title={Low-energy spectrum of double-junction superconducting circuits in the Born-Oppenheimer approximation}, 
      author={Leo Uhre Jakobsen and Ksenia Shagalov and David Feldstein-Bofill and Morten Kjaergaard and Karsten Flensberg and Svend Krøjer},
      year={2026},
      eprint={2603.26374},
      archivePrefix={arXiv},
      primaryClass={quant-ph},
      url={https://arxiv.org/abs/2603.26374}, 
}

@article{Zorin2006,
  title = {Bloch Inductance in Small-Capacitance Josephson Junctions},
  author = {Zorin, A. B.},
  journal = {Phys. Rev. Lett.},
  volume = {96},
  issue = {16},
  pages = {167001},
  numpages = {4},
  year = {2006},
  month = {Apr},
  publisher = {American Physical Society},
  doi = {10.1103/PhysRevLett.96.167001},
  url = {https://link.aps.org/doi/10.1103/PhysRevLett.96.167001}
}

@phdthesis{Cailleaux2025,
  title={Driven-dissipative Josephson dynamics in a high-impedance many-body environment},
  author={Cailleaux, Samuel},
  year={2025},
  school={Universit{\'e} Grenoble Alpes}
}

@article{Koch2009,
  title = {Charging Effects in the Inductively Shunted Josephson Junction},
  author = {Koch, Jens and Manucharyan, V. and Devoret, M. H. and Glazman, L. I.},
  journal = {Phys. Rev. Lett.},
  volume = {103},
  issue = {21},
  pages = {217004},
  numpages = {4},
  year = {2009},
  month = {Nov},
  publisher = {American Physical Society},
  doi = {10.1103/PhysRevLett.103.217004},
  url = {https://link.aps.org/doi/10.1103/PhysRevLett.103.217004}
}

@book{Landau1980,
  title     = {Statistical Physics: Theory of the Condensed State},
  author    = {Landau, L. D. and Lifshitz, E. M. and Pitaevskii, L. P.},
  year      = {1980},
  volume    = {9},
  publisher = {Butterworth-Heinemann}
  }

@article{Pechenezhskiy2020,
  title={The superconducting quasicharge qubit},
  author={Pechenezhskiy, Ivan V and Mencia, Raymond A and Nguyen, Long B and Lin, Yen-Hsiang and Manucharyan, Vladimir E},
  journal={Nature},
  volume={585},
  number={7825},
  pages={368--371},
  year={2020},
  publisher={Nature Publishing Group UK London}
}

@book{Magnus2013,
  title={Hill's Equation},
  author={Magnus, W. and Winkler, S.},
  isbn={9780486150291},
  series={Dover phoenix editions},
  url={https://books.google.fr/books?id=XwEMurjYXvYC},
  year={2013},
  publisher={Dover Publications}
}

@article{Rymarz2021,
  title = {Hardware-Encoding Grid States in a Nonreciprocal Superconducting Circuit},
  author = {Rymarz, Martin and Bosco, Stefano and Ciani, Alessandro and DiVincenzo, David P.},
  journal = {Phys. Rev. X},
  volume = {11},
  issue = {1},
  pages = {011032},
  numpages = {25},
  year = {2021},
  month = {Feb},
  publisher = {American Physical Society},
  doi = {10.1103/PhysRevX.11.011032},
  url = {https://link.aps.org/doi/10.1103/PhysRevX.11.011032}
}

@article{Morse1929,
  title = {Diatomic Molecules According to the Wave Mechanics. II. Vibrational Levels},
  author = {Morse, Philip M.},
  journal = {Phys. Rev.},
  volume = {34},
  issue = {1},
  pages = {57--64},
  numpages = {0},
  year = {1929},
  month = {Jul},
  publisher = {American Physical Society},
  doi = {10.1103/PhysRev.34.57},
  url = {https://link.aps.org/doi/10.1103/PhysRev.34.57}
}

@article{BO1927,
author = {Born, M. and Oppenheimer, R.},
title = {Zur Quantentheorie der Molekeln},
journal = {Annalen der Physik},
volume = {389},
number = {20},
pages = {457-484},
doi = {https://doi.org/10.1002/andp.19273892002},
year = {1927}
}

@article{Orlikowski1974,
title = {Nonadiabatic Corrections To The Vibrational Energies Of The Hydrogen Molecule},
journal = {Chemical Physics Letters},
volume = {24},
number = {4},
pages = {461-463},
year = {1974},
issn = {0009-2614},
doi = {https://doi.org/10.1016/0009-2614(74)80156-9},
url = {https://www.sciencedirect.com/science/article/pii/0009261474801569},
author = {T. Orlikowski and L. Wolniewicz},
}

@article{Schrieffer1966,
  title = {Relation between the Anderson and Kondo Hamiltonians},
  author = {Schrieffer, J. R. and Wolff, P. A.},
  journal = {Phys. Rev.},
  volume = {149},
  issue = {2},
  pages = {491--492},
  numpages = {0},
  year = {1966},
  month = {Sep},
  publisher = {American Physical Society},
  doi = {10.1103/PhysRev.149.491},
  url = {https://link.aps.org/doi/10.1103/PhysRev.149.491}
}

@article{Luttinger1955,
  title = {Motion of Electrons and Holes in Perturbed Periodic Fields},
  author = {Luttinger, J. M. and Kohn, W.},
  journal = {Phys. Rev.},
  volume = {97},
  issue = {4},
  pages = {869--883},
  numpages = {0},
  year = {1955},
  month = {Feb},
  publisher = {American Physical Society},
  doi = {10.1103/PhysRev.97.869},
  url = {https://link.aps.org/doi/10.1103/PhysRev.97.869}
}

@article{Zinnjustin2004,
title = {Multi-instantons and exact results II: specific cases, higher-order effects, and numerical calculations},
journal = {Annals of Physics},
volume = {313},
number = {2},
pages = {269-325},
year = {2004},
issn = {0003-4916},
doi = {https://doi.org/10.1016/j.aop.2004.04.003},
url = {https://www.sciencedirect.com/science/article/pii/S0003491604000673},
author = {Jean Zinn-Justin and Ulrich D. Jentschura},
}

@InProceedings{Dunne2017,
author="Dunne, Gerald V.
and {\"U}nsal, Mithat",
editor="Fauvet, Fr{\'e}d{\'e}ric
and Manchon, Dominique
and Marmi, Stefano
and Sauzin, David",
title="WKB and resurgence in the Mathieu equation",
booktitle="Resurgence, Physics and Numbers",
year="2017",
publisher="Scuola Normale Superiore",
address="Pisa",
pages="249--298",
isbn="978-88-7642-613-1"
}

@article{Vakhtel2024,
  title = {Tunneling of fluxons via a Josephson resonant level},
  author = {Vakhtel, T. and Kurilovich, P. D. and Pita-Vidal, M. and Bargerbos, A. and Fatemi, V. and van Heck, B.},
  journal = {Phys. Rev. B},
  volume = {110},
  issue = {4},
  pages = {045404},
  numpages = {10},
  year = {2024},
  month = {Jul},
  publisher = {American Physical Society},
  doi = {10.1103/PhysRevB.110.045404},
  url = {https://link.aps.org/doi/10.1103/PhysRevB.110.045404}
}

@article{Devoret2021DoesBJ,
  title={Does Brian Josephson’s Gauge-Invariant Phase Difference Live on a Line or a Circle?},
  author={Michel H. Devoret},
  journal={Journal of Superconductivity and Novel Magnetism},
  year={2021},
  volume={34},
  pages={1633 - 1642},
  url={https://api.semanticscholar.org/CorpusID:236346880}
}

@article{Ithier2005,
  title = {Decoherence in a superconducting quantum bit circuit},
  author = {Ithier, G. and Collin, E. and Joyez, P. and Meeson, P. J. and Vion, D. and Esteve, D. and Chiarello, F. and Shnirman, A. and Makhlin, Y. and Schriefl, J. and Sch\"on, G.},
  journal = {Phys. Rev. B},
  volume = {72},
  issue = {13},
  pages = {134519},
  numpages = {22},
  year = {2005},
  month = {Oct},
  publisher = {American Physical Society},
  doi = {10.1103/PhysRevB.72.134519},
  url = {https://link.aps.org/doi/10.1103/PhysRevB.72.134519}
}

@article{Matveev2002,
  title = {Persistent Current in Superconducting Nanorings},
  author = {Matveev, K. A. and Larkin, A. I. and Glazman, L. I.},
  journal = {Phys. Rev. Lett.},
  volume = {89},
  issue = {9},
  pages = {096802},
  numpages = {4},
  year = {2002},
  month = {Aug},
  publisher = {American Physical Society},
  doi = {10.1103/PhysRevLett.89.096802},
  url = {https://link.aps.org/doi/10.1103/PhysRevLett.89.096802}
  }
\clearpage
\setcounter{section}{0}
\renewcommand{\thesection}{Appendix \Alph{section}}
\renewcommand{\thesubsection}{\arabic{subsection}}

\section{Multi-band Born-Oppenheimer model}
\label{sec:multiBO}
Here we derive the exact formulation of the circuit Hamiltonian \eqref{eq:Hfull} in the Born-Oppenheimer framework.

Since $\hat N = -i \frac{d}{d\varphi}$, we rewrite the Hamiltonian as
\begin{equation}
    {\hat H}_\mathrm{full} = - 4\varepsilon E_{C_J} \frac{d^2}{d \varphi^2} + {\hat H}_\mathrm{light}( \hat \varphi, \phiext)
\end{equation}
where $ \displaystyle \varepsilon = \frac{E_{C_S}}{E_{C_J}}$ is the small parameter controlling the energy scale separation.

We recall that ${\hat H}_\mathrm{light}$ is the Hamiltonian of two uncoupled fluxonia; its eigenstates decompose as $\ket{\Psi_{mn}^\mathrm{light}}(\varphi, \phiext) = \ket{\psi_m^L}(\varphi) \otimes \ket{\psi_n^R}(\varphi - \phiext)$ for $m,n \in \mathbb{N}$ and their corresponding eigenvalues $E^\mathrm{light}_{mn}(\varphi, \phiext) = E_n^L + E_n^R$. We consider the basis of the full Hilbert space spanned by
$$ \ket{\varphi,m,n} =  \ket{\varphi} \otimes \ket{\Psi_{mn}^\mathrm{light}}(\varphi, \phiext)\;,$$ where
$\varphi \in \mathbb{R}, \ (m,n)\in \mathbb{N}^2\;.$
Any vector $\ket{\Psi}$ can then be decomposed as
$$ \ket{\Psi} = \sum_{m,n} \int \mathrm{d}\varphi \ \chi_{mn}(\varphi) \ket{\varphi,m,n} \ ,$$
with $\chi_{mn}(\varphi)$ the slow-mode wavefunction corresponding to the $(m,n)$ band of the fast modes. The action of ${\hat H}_\mathrm{full}$ on $\ket{\Psi}$ writes as
\begin{align}
    {\hat H}_\mathrm{full} \ket{\Psi} =  \sum_{m,n}& \int \mathrm{d}\varphi \ket{\varphi}\otimes \notag \\
    \bigg(  &- 4\varepsilon E_{C_J} \frac{d^2}{d\varphi^2}\Big(\chi_{mn}(\varphi) \ket{\Psi_{mn}^\mathrm{light}}(\varphi, \phiext)\Big)\notag \\
    &+ E_{mn}^\mathrm{light}(\varphi ,\phiext)\chi_{mn}(\varphi) \ket{\Psi_{mn}^\mathrm{light}}(\varphi, \phiext)\bigg)
\end{align}
Integrating against $\bra{\varphi} \otimes \bra{\Psi_{kl}^\mathrm{light}}(\varphi ,\phiext) $, the time-independent Schrödinger equation ${\hat H}_\mathrm{full} \ket{\Psi} = E \ket{\Psi}$ takes the form
\begin{align}
    E \chi_{kl}(\varphi) =  &- 4\varepsilon E_{C_J} \frac{d^2\chi_{kl}}{d\varphi^2} + E_{kl}^\mathrm{light}(\varphi ,\phiext)\chi_{kl}(\varphi)\notag \\
    &\quad- 4\varepsilon E_{C_J}\sum_{m,n} \begin{aligned}[t] \Big(&2A_{kl,mn}(\varphi, \phiext) \frac{d\chi_{mn}}{d\varphi} \\
    &+ B_{kl,mn}(\varphi, \phiext) \chi_{mn}(\varphi) \Big) \ ,\end{aligned}
    \label{eigenenergy_eq}
\end{align}

where 
\begin{equation}
    A_{kl,mn}(\varphi, \phiext) = \braket{\Psi_{kl}^\mathrm{light}} {\partial_\varphi \Psi_{mn}^\mathrm{light}}(\varphi, \phiext)
\end{equation}
and
\begin{equation}
    B_{kl,mn}(\varphi, \phiext) = \braket{\Psi_{kl}^\mathrm{light}} {\partial_\varphi^2 \Psi_{mn}^\mathrm{light}}(\varphi, \phiext) \;.
\end{equation}
\begin{figure}
    \centering
    \includegraphics[scale=1]{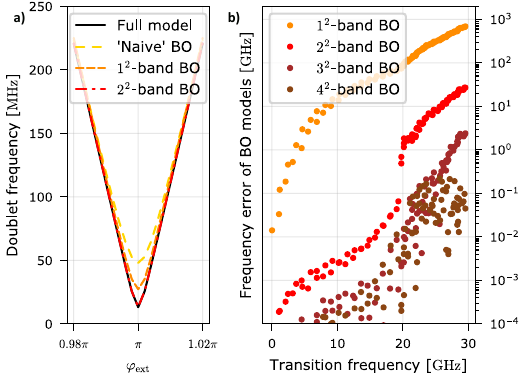}
    \caption{(a) Doublet transition frequency (y-axis) versus external flux (x-axis) close to the sweet spot, computed from the full lumped model (black), the ‘naive' Born-Oppenheimer Hamiltonian omitting the geometric potential (yellow), the 1-band model (orange) and the $2^2$-band model (red). The naive model predicts too large a frequency at half-flux by a factor 4, showcasing the importance of the geometric potential and the higher-band contributions. (b) Discrepancy of predicted transition frequencies (y-axis) between the full lumped model and BO models for various band cutoffs versus transition frequency in the full lumped model (x-axis). The $2^2$-band model is sufficient to capture the spectrum up to 20 GHz with an error below 100 MHz, corresponding to a relative error of $3\%$ over the entire frequency range.}
    \label{fig:multiband_bo}
\end{figure}
Note that all the terms in Eq.~\eqref{eigenenergy_eq} are $2\pi$-periodic in $\varphi$; the slow mode can therefore be described by a discrete charge variable $\hat N$ and an offset charge $N_g$, such that $\chi_{kl}(\varphi + 2\pi) = \mathrm{e}^{2i\pi N_g} \chi_{kl}(\varphi)$.
The Born-Oppenheimer Hamiltonian is then composed of blocks $\hat H^{BO}_{kl, mn}$, each acting on the slow variable:
\begin{align}
    H^{BO}_{kl, mn} = &\delta_{km}\delta_{ln} \Big(  4\varepsilon E_{C_J} (\hat N-N_g)^2  + E_{mn}^\mathrm{light}(\hat\varphi ,\phiext) \Big) \notag \\
    &-4\varepsilon E_{C_J} \big( 2i A_{kl,mn}(\hat \varphi, \phiext) (\hat N - N_g) \notag \\
    &+ B_{kl,mn}(\hat \varphi, \phiext) \big) \ .
    \label{BO_multiband}
\end{align}
For any $k,l \in \mathbb{N}$, $ \displaystyle A_{kl,kl} = \frac{1}{2}\partial_\varphi \rvert\rvert \Psi_{kl}^\mathrm{fast} \rvert \rvert^2 = 0$ since the fluxonium wavefunctions can be chosen to be real-valued.

Moreover, since the fast dynamics can be separated into two independent modes, we have
\begin{align}
    A_{kl,mn}(\varphi, \phiext) = &\delta_{ln}\braket{\psi_{k}^L} {\partial_\varphi \psi_{m}^L} (\varphi) \notag \\
    &+ \delta_{km}\braket{\psi_{l}^R} {\partial_\varphi \psi_{n}^R} (\varphi - \phiext)
\end{align}
and
\begin{align}
    B_{kl,mn}(\varphi, \phiext) = &\delta_{ln}\braket{\psi_{k}^L} {\partial_\varphi^2 \psi_{m}^L} (\varphi) \notag \\
    &+ \delta_{km}\braket{\psi_{l}^R} {\partial_\varphi^2 \psi_{n}^R} (\varphi - \phiext) \notag \\
    &+ 2 \braket{\psi_{k}^L} {\partial_\varphi \psi_{m}^L} (\varphi)\braket{\psi_{l}^R} {\partial_\varphi \psi_{n}^R} (\varphi - \phiext)
\end{align}
Approximate solutions can be obtained by truncating up to any number of bands; Eq. \eqref{H_BO} in the main text is obtained by keeping only the fundamental band
\begin{align}
        H^{BO}_{00, 00} =  4&\varepsilon E_{C_J} \hat (\hat N-N_g)^2  + E_{00}^\mathrm{light}(\hat\varphi ,\phiext) \notag \\
    &-4\varepsilon E_{C_J} B_{00,00}(\hat \varphi, \phiext)
\end{align}

In practice, numerical diagonalization of the fluxonium Hamiltonian is carried out in the gauge where the external flux enters in the cosine term -- unlike the formulation of $ \hat H_\mathrm{fluxo}^{L/R}$ in Eq.~\ref{eq:Hfluxo} :
\begin{equation}
    \tilde H_\mathrm{fluxo}^{L/R} (\alpha) = 4 E_{CJ} \hat n_{L/R}^2
- E_J \cos(\hat \theta_{L/R}+\alpha)+\frac{1}{2}E_L  \hat \theta_{L/R} ^2 \ ,
\end{equation}
yielding eigenstates $\ket{\tilde \psi_n^{L/R}}(\alpha)$. One may connect these two gauges choices through the unitary operator $\hat U = \mathrm{e}^{-i\alpha \hat n_{L/R}}$ :
\begin{equation}
    \ket{ \psi_n^{L/R}}(\alpha) = \hat U \ket{\tilde \psi_n^{L/R}}(\alpha) = \mathrm{e}^{-i\alpha \hat n_{L/R}}\ket{\tilde \psi_n^{L/R}}(\alpha) \ .
\end{equation}
Therefore, the coefficients $A_{kl,mn}$ are computed with the formula
\begin{align}
    A_{kl,mn}(\varphi, \phiext) = &\delta_{ln}\braket{\tilde\psi_{k}^L} {\partial_\varphi \tilde\psi_{m}^L} (\varphi) \notag \\
    & -i\delta_{ln}\bra{\tilde\psi_{k}^L}\hat  n_L\ket{\tilde\psi_{m}^L} (\varphi) \notag \\
    &+ \delta_{km}\braket{\tilde \psi_{l}^R} {\partial_\varphi \tilde\psi_{n}^R} (\varphi - \phiext) \notag \\
    & -i\delta_{km}\bra{\tilde\psi_{l}^R}\hat  n_R\ket{\tilde\psi_{n}^R} (\varphi - \phiext) \ ,
\end{align}
where we used $\partial_\varphi \hat U = -i \hat n_{L/R} \hat U$, and the $B_{kl,mn}$ coefficients are obtained through the completeness relation
\begin{align}
    \braket{\psi_{m}^{L/R}} {\partial_\varphi^2 \psi_{n}^{L/R}} &= \partial_\varphi \braket{\psi_{m}^{L/R}} {\partial_\varphi \psi_{n}^{L/R}} \notag \\
    &+ \sum_k \braket{\psi_{m}^{L/R}} {\partial_\varphi \psi_{k}^{L/R}} \braket{\psi_{k}^{L/R}} {\partial_\varphi \psi_{n}^{L/R}}\ .
\end{align}
\section{Effective single-mode models}
\subsection{Schrieffer-Wolff transformation}
Although the multi-band Born-Oppenheimer model offers an increased degree of precision for the qubit transition frequencies, its interpretation is less straightforward due to the multi-component nature of the slow wavefunction.

The Schrieffer-Wolff (SW) transformation \cite{Luttinger1955, Schrieffer1966} precisely enacts a unitary transformation that reframes the effect of virtual transitions to higher bands and back into an effective single-mode Hamiltonian. In this section we present the 2nd-order Schrieffer-Wolff Hamiltonian obtained from \eqref{BO_multiband}.

The transformed Hamiltonian is defined as 
\begin{equation}
    \mathcal{\hat H}_{SW} = \mathrm{e}^{\varepsilon \hat S} \mathcal{\hat H} \mathrm{e}^{-\varepsilon \hat S} = \mathcal{\hat H} + \varepsilon[\hat S, \mathcal{\hat H}] + \frac{\varepsilon^2}{2} [\hat S, [\hat S, \mathcal{\hat H}]] + \mathcal{O}(\varepsilon^3)
\end{equation}
where $\hat S = - \hat S^\dagger$ is the generator of the transformation. We look for $\hat S$ such that, up to order $\varepsilon$, the transformed Hamiltonian has no off-diagonal blocks coupling the fundamental band $(0,0)$ to higher-energy ones. Defining the orthogonal projectors
\begin{align}
    \mathcal{\hat P} &= \int \mathrm{d}\varphi \ket{\varphi}\otimes \ket{\Psi_{00}^\mathrm{light}}(\hat\varphi ,\phiext) \bra{\Psi_{00}^\mathrm{light}}(\hat\varphi ,\phiext) \otimes \bra{\varphi} \notag \\
    \mathcal{\hat Q} &= 1- \mathcal{\hat P} \ ,
\end{align}
the condition for $\hat S$ writes as
\begin{equation}
    \mathcal{\hat P} \left(\varepsilon[\hat S, \mathcal{\hat H}] + \mathcal{\hat H}  \right) \mathcal{\hat Q} = \mathcal{O}(\varepsilon^2) \ .
\end{equation}
The only off-block-diagonal contribution to $\mathcal{\hat H}$ is the kinetic term $4\varepsilon E_{C_J} \hat N^2$, which yields at order $\varepsilon$ for all bands $(m,n)$ :
\begin{align}
    4 E_{C_J} \hat N^2\rvert_{00,mn} &= \bra{\Psi_{00}^\mathrm{light}} [ {\hat H}_\mathrm{light}, \hat S] \ket{\Psi_{mn}^\mathrm{light}}(\hat\varphi ,\phiext) \notag \\
    & \ \begin{aligned}
        =E_{00}^\mathrm{light}&(\hat \varphi ,\phiext) \hat S\rvert_{00,mn} \\
        &-  \hat S\rvert_{00,mn} E_{m,n}^\mathrm{light}(\hat \varphi ,\phiext) \end{aligned}
\end{align}
where the subscripts $\rvert_{00,mn}$ denote the projection of the operators $\hat S, \hat N^2$ on block $(00,mn)$, and
\begin{equation}
    \hat N^2\rvert_{00,mn} = - 2i A_{00,mn}(\hat \varphi, \phiext) \hat N - B_{00,mn}(\hat \varphi, \phiext)
\end{equation}
from \eqref{BO_multiband}.
The solution of this Sylvester equation takes the form
\begin{align}
    S\rvert_{00,mn} = 4 E_{C_J} \bigg(&\frac{  2i A_{00,mn}(\hat \varphi, \phiext) \hat N + B_{00,mn}(\hat \varphi, \phiext)}{E_{mn}^\mathrm{light}(\hat \varphi ,\phiext) -E_{00}^\mathrm{light}(\hat \varphi ,\phiext)} \notag \\
    &- \frac{  2 A_{00,mn}(\hat \varphi, \phiext) \partial_\varphi E_{mn}^\mathrm{light}(\hat \varphi ,\phiext)}{\left( E_{mn}^\mathrm{light}(\hat \varphi ,\phiext) -E_{00}^\mathrm{light}(\hat \varphi ,\phiext)\right)^2}\bigg) \ .
    \label{S_00mn}
\end{align}
Similarly,
\begin{align}
    S\rvert_{mn,00} = -4 E_{C_J} \bigg(&\frac{  2i A_{mn,00}(\hat \varphi, \phiext) \hat N + B_{mn,00}(\hat \varphi, \phiext)}{E_{mn}^\mathrm{light}(\hat \varphi ,\phiext) -E_{00}^\mathrm{light}(\hat \varphi ,\phiext)} \notag \\
    &+ \frac{  2 A_{mn,00}(\hat \varphi, \phiext) \partial_\varphi E_{0,0}^\mathrm{light}(\hat \varphi ,\phiext)}{\left( E_{mn}^\mathrm{light}(\hat \varphi ,\phiext) -E_{00}^\mathrm{light}(\hat \varphi ,\phiext)\right)^2}\bigg) \ .
    \label{S_mn00}
\end{align}
Since the transformed Hamiltonian has no coupling between the $\mathcal{\hat P}$ and $\mathcal{\hat Q}$ sectors, the low-energy dynamics is completely encompassed at order $\varepsilon^2$ by
\begin{align}
    \mathcal{\hat H}_{SW}\rvert_{00,00} = H^{BO}_{00, 00} &+ \varepsilon \bra{\Psi_{00}^\mathrm{light}}[\hat S, \mathcal{\hat H}] \ket{\Psi_{00}^\mathrm{light}}\notag \\
    &+ \frac{\varepsilon^2}{2} \bra{\Psi_{00}^\mathrm{light}}  [\hat S, [\hat S, \mathcal{\hat H}]] \ket{\Psi_{00}^\mathrm{light}} \ .
\end{align}
Using the expressions \eqref{S_00mn} and \eqref{S_mn00}, we finally obtain the effective Hamiltonian
\begin{align}
    \mathcal{\hat H}_{SW}\rvert_{00,00} = &4\hat N E_C^\mathrm{eff} (\hat \varphi, \phiext)\hat N \notag \\
    &+ E_{00}^\mathrm{light}(\hat\varphi ,\phiext) 
    -4\varepsilon E_{C_J} B_{00,00}(\hat \varphi, \phiext)\notag \\
    &+ \hat U_\mathrm{SW}(\hat \varphi, \phiext) \ ,
    \label{eq:H_SW}
\end{align}
where
\begin{equation}
  E_C^\mathrm{eff} =  \varepsilon E_{C_J} \left(1 - 16 \varepsilon E_{C_J} \sum_{(m,n)\neq(0,0)} \frac{\lvert A_{mn,00} \rvert^2 }{E_{mn}^\mathrm{light} -E_{00}^\mathrm{light}} \right) 
  \label{ECeff}
\end{equation}
and
\begin{align}
    \hat U_\mathrm{SW} = 16\varepsilon^2 &E_{C_J}^2 \sum_{(m,n)\neq(0,0)} \bigg(-\frac{B_{00,mn} B_{mn,00}}{E_{mn}^\mathrm{light} -E_{00}^\mathrm{light}} \notag \\
    &- \frac{2A_{00,mn} \partial_\varphi B_{mn,00}}{E_{mn}^\mathrm{light} -E_{00}^\mathrm{light}} \notag \\
        &- \frac{2A_{mn,00}  B_{00,mn}\partial_\varphi E_{0,0}^\mathrm{light}}{\big( E_{mn}^\mathrm{light} -E_{00}^\mathrm{light}\big)^2 } \notag   \\
         &+ \frac{2A_{00,mn}  B_{mn,00}\partial_\varphi (E_{mn}^\mathrm{light}-E_{00}^\mathrm{light})}{\big( E_{mn}^\mathrm{fast} -E_{00}^\mathrm{light}\big)^2 } \notag   \\
         & + 2 \lvert A_{00,mn} \rvert^2 \partial_\varphi \left( \frac{\partial_\varphi E_{00}^\mathrm{light}}{\big( E_{mn}^\mathrm{light} -E_{00}^\mathrm{light}\big)^2} \right) \bigg)
         \label{Usw}
\end{align}

\begin{figure}
    \centering
    \includegraphics[scale=1]{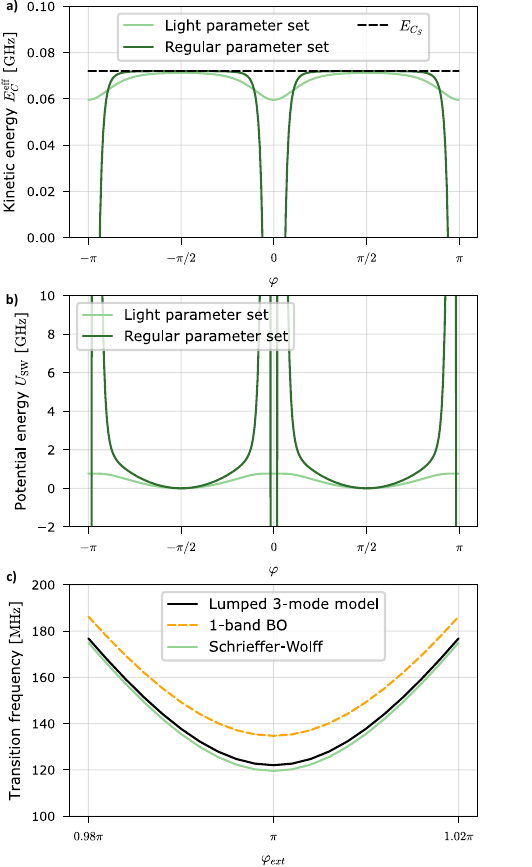}
    \caption{(a) Effective kinetic energy $E_C^\mathrm{eff}$ (y-axis) versus $\varphi$ (x-axis), computed for the parameter set of Table~\ref{table1} (dark green) and a lighter parameter set $(E_J/3, E_{C_J}, E_L, E_{C_S})$ (light green). The physical shunt capacitive energy $E_{C_S}$ is plotted in black dashes. (b) Effective potential $U_{SW}$ (y-axis) versus $\varphi$ (x-axis) for the two same parameter sets. In both panels, the regular parameter set (dark green) features sharp divergences, signalling a breakdown of the Schrieffer-Wolff perturbative expansion. (c) Doublet transition frequency (y-axis) versus external flux around $\pi$ (x-axis), computed with the full lumped (solid black line), the single-band BO (dashed orange line) and SW (solid light green line) with the light parameter set, showing increased quantitative agreement from the inclusion of second-order SW corrections. The results for the regular parameter set are not presented, due to the SW Hamiltonian's numerical instability in that regime.}
    \label{fig:sw}
\end{figure}
Fig. \ref{fig:sw} a,b shows the effective kinetic energy and potential entering the transformedy Hamiltonian, for the parameter set of Table \ref{table1} as well as one with lighter junctions of Josephson energy $E_J/3$ and capacitive energy $E_{C_J}$. The light parameter set features a renormalization of the kinetic energy $E_C^\mathrm{eff} (\hat \varphi, \phiext = \pi) \leq E_{C_S}$, due to the effect of virtual transitions to higher bands and back. The obtained doublet transition frequencies in the vicinity of $\phiext = \pi$ (Fig. \ref{fig:sw} c) show improved agreement with the lumped model compared to the single-band Born-Oppenheimer Hamiltonian.

In contrast, the effective kinetic energy obtained from the regular parameter set shows a sign change around $\varphi=0$ and $\pi$, signaling a breakdown of the perturbative Schrieffer-Wolff expansion. This leads to numerical instabilities in the Hamiltonian and precludes its practical diagonalization. It follows from equations \eqref{ECeff} and \eqref{Usw} that this breakdown occurs when $E_{C_S}$ becomes comparable with $E_{m,n}^\mathrm{fast} -E_{0,0}^\mathrm{fast}$, i.e. when $\varepsilon_\mathrm{ps} \lesssim E_{C_S}$ ; we define this threshold as the onset of the multi-mode physics region (Fig. \ref{fig:fig5}).
\subsection{Linear response}
\begin{figure*}[ht]
    \includegraphics[scale=1]{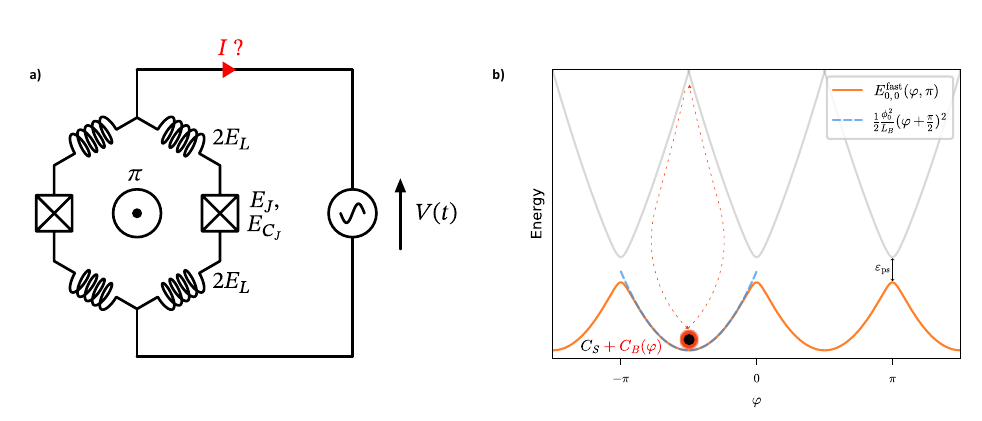}
    \caption{(a) Circuit diagram of the KITE dipole, biased by an alternating voltage source. The linear response to the pertubation $V(t)$ is encoded in the current across the dipole $I$. (b) Schematic representation of the capacitive loading effect of higher bands. Within the linear response framework, an effective particle located at the bottom of the fundamental band's well feels a confining potential given by the band's curvature; its inertial term is composed of the shunt capacitance $C_S$ and the Bloch capacitance $C_B(\varphi)$, arising from virtual transitions to higher excited bands and back. The strength of this effect is controlled by the interband gap, given by the phase-slip energy $\varepsilon_\mathrm{ps}$.}
    \label{fig:linear_response}
\end{figure*}

We now give a complementary perspective on the capacitive loading observed in the Schrieffer-Wolff Hamiltonian. Taking the dual of the method presented in \cite{Zorin2006, Cailleaux2025} to determine the impedance of an isolated Josephson junction, we examine the linear response of the KITE dipole subjected to a small voltage bias $V(t)$ (Fig. \ref{fig:linear_response}a).

The interband coupling operator introduced in Eq.\eqref{Nop} has the following matrix elements \cite{Koch2009}:
\begin{equation}
    \bra{\varphi, k,l} \mathcal{N}^\mathrm{interband} \ket{\varphi',m,n}= -i \delta(\varphi-\varphi') A_{kl,mn}(\varphi, \phiext)\ ;
\end{equation}
it is diagonal in $\varphi$.
The current operator $\hat I$ is found as the time derivative of $-2e\hat N$:
\begin{equation}
    \hat I = -\frac{2e}{i\hbar} [\hat N, {\hat H}_\mathrm{light}(\hat  \varphi, \phiext)] \ ,
    \label{currentop}
\end{equation}
and the current variation across the dipole due to the voltage bias is given by the Kubo formula:
\begin{equation}
    \langle \Delta \hat I(t) \rangle = - \frac{1}{\hbar} \int_{-\infty}^{+\infty} \mathrm{d}\tau \ G^r_{IN}(t-\tau)  2e V(\tau) \ ,
\end{equation}
where 
\begin{equation}
G^r_{IN}(t) = i \theta(t) \langle [\hat I(t), \hat N(0)]
\label{greensfunction}
\end{equation}
is the retarded Green's function, $\theta(t)$ is the Heaviside distribution, and expectation values are taken over $\ket{\varphi,0,0}$.

The system's linear response can then be expressed in frequency space in terms of the system's admittance:
\begin{equation}
    Y(\omega) = - \frac{1}{\phi_0} G^r_{IN}(\omega) \ .
\end{equation}
Differentiating Eq. \eqref{greensfunction} yields
\begin{equation}
    \partial_t G^r_{IN}(t) = i \delta(t) \langle [\hat I, \hat N] \rangle + \frac{1}{2e} G^r_{II}(t) \ ,
\end{equation}
and so
\begin{equation}
    i \phi_0 \omega Y(\omega) = i  \langle [\hat I, \hat N] \rangle +  \frac{1}{2e} G^r_{II}(\omega) \ .
\end{equation}
From Eqs. \eqref{Nop} and \eqref{currentop}, we can compute the first term
\begin{equation}
    \langle [\hat I, \hat N] \rangle = -\frac{2e}{i\hbar} \left(  \partial_\varphi^2 E_{00}^\mathrm{light} + 2\sum_{(m,n)\neq(0,0)} \Delta_{mn} \lvert \mathcal{N}_{mn}\rvert^2 \right)
\end{equation}
where we defined $\Delta_{mn} =  E_{mn}^\mathrm{light} -  E_{00}^\mathrm{light}$ and
 $\mathcal{N}_{mn} =\bra{m,n}\mathcal{\hat N}^\mathrm{interband}\ket{0,0}$.
 
 We then write the second term
 \begin{align}
     G_{II}^r(\omega) &= \sum_{(m,n)\neq(0,0)} \lvert \bra{m,n} \hat I \ket{0,0}\rvert ^2 \left( \frac{1}{\omega + \frac{\Delta_{mn}}{\hbar}} - \frac{1}{\omega - \frac{\Delta_{mn}}{\hbar}}\right) \notag \\
     &= -\frac{4e^2}{\hbar} \sum_{(m,n)\neq(0,0)} \lvert  \mathcal{N}_{mn}\rvert^2 \frac{2 \Delta_{mn}^3}{\hbar^2 \omega^2 - \Delta_{mn}^2} \ .
 \end{align}

 Finally, we obtain the expression for the dipole's admittance:
 \begin{equation}
     Y(\omega) = -\frac{1}{i\omega \phi_0^2}  \partial_\varphi^2 E_{00}^\mathrm{light} + 8ie^2  \sum_{(m,n)\neq(0,0)} \lvert  \mathcal{N}_{mn}\rvert^2 \frac{\omega \Delta_{mn}}{\hbar^2\omega^2- \Delta_{mn}^2} \ .
 \end{equation}
 We focus on the low-energy regime, i.e. $\hbar \omega \ll \Delta_{mn}$, which is relevant to investigate the response around the doublet transition frequency. The admittance then takes the simpler form
 \begin{equation}
     Y(\omega) = -\frac{1}{i\omega L_B(\varphi)} - i \omega C_B(\varphi) \ ,
 \end{equation}
 with
 \begin{align}
     L_B(\varphi) &= \frac{\phi_0^2}{\partial_\varphi^2 E_{00}^\mathrm{light}}\ , \\
     C_B(\varphi) &= 8e^2 \sum_{(m,n)\neq(0,0)} \frac{\lvert  \mathcal{N}_{mn}\rvert^2}{\Delta_{mn}} \ .
 \end{align}

 The Bloch inductance $L_B$ simply corresponds to the curvature of the base band potential (not capturing the geometric potential of Eq. \eqref{H_BO}), while the Bloch capacitance $C_B$ accounts for the capacitive loading of the dipole due to virtual transitions to higher bands and back (see Fig.\ref{fig:linear_response} b).

 Taking a step back to consider the full circuit, the total capacitive energy is written as
 \begin{align}
     \frac{e^2}{2(C_S + C_B)} &= E_{C_S} \left( 1 - \frac{C_B}{C_S} + \mathcal{O}((C_B/C_S)^2) \right) \notag \\
     &= E_{C_S} \left( 1 - 16 E_{C_S} \sum_{(m,n)\neq(0,0)} \frac{\lvert  \mathcal{N}_{mn}\rvert^2}{\Delta_{mn}}\right)
 \end{align}
 which exactly matches, up to order $\varepsilon^2$, the expression given in \eqref{ECeff} for the effective kinetic energy of the Schrieffer-Wolff-transformed Hamiltonian.
\section{Blochnium regime for the KITE transmon}
\label{sec:blochnium}
We have shown in Fig.\ref{fig:fig4} that the selection rules for the $\cos 2\varphi$ transmon still allow transitions between the logical states $(\ket{0+}, \ket{0-})$ through the $\hat \theta_\Delta$ and $\hat n_\Delta$ operators. However, there exists a regime of parameters where these matrix elements become vanishingly small, thereby providing a fully $T_1$-protected qubit.
\begin{figure}
    \centering
    \includegraphics[scale=1]{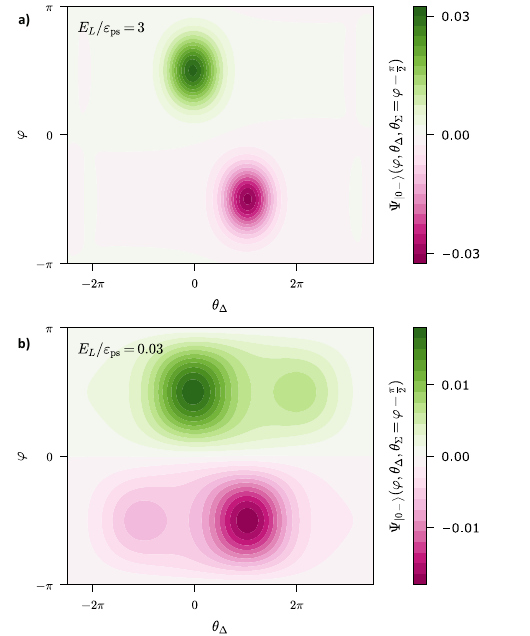}
    \caption{Wavefunction cuts of state $\ket{0-}$ along the plane $\theta_\Sigma=\varphi - \frac{\pi}{2}$, with parameter sets (a): $(E_J, E_{C_J}, E_L, E_{C_S})$ (see Table \ref{table1}) and (b): $(E_J/2, E_{C_J}\times 2, E_L/10, E_{C_S})$}
    \label{fig:blochnium}
\end{figure}

In the regime of Table \ref{table1}, the ground state doublet consists of symmetric and antisymmetric superpositions of two wells located at $(\varphi, \theta_L, \theta_R)= (-\frac{\pi}{2},0,-2\pi)$ and $(\frac{\pi}{2},0,0)$ (Fig. \ref{fig:blochnium}a). But when the fluxonium degrees of freedom $\theta_L$, $\theta_R$ enter the blochnium regime \cite{Pechenezhskiy2020}, defined as $E_L \ll \varepsilon_\mathrm{ps}$, their wavefunctions delocalize over several potential wells, leading to the interleaving pattern seen in Fig.\ref{fig:blochnium}b.

In turn, the $\hat \theta_\Delta$ matrix element is suppressed, going from $1.38$ to $0.126$ between the two parameter sets, corresponding respectively to $E_L/\varepsilon_\mathrm{ps} = 3$ and $0.03$.
Note that this asymptotic $T_1$ protection is a multimode feature that could not be obtained with a single-mode circuit \cite{GyenisReview}: in particular, the blochnium qubit still has a large matrix element between its two logical states, regardless of the parameter regime.

However, this asymptotic regime, while satisfactory from the point of view of matrix elements, is undesirable for qubit protection. Indeed, as the ratio $E_L/\varepsilon_\mathrm{ps}$ diminishes and the fluxonium degrees of freedom enter the blochnium regime, they become insensitive to external flux \cite{Pechenezhskiy2020} and thus the confining potential amplitude $E_{J2}$ is exponentially suppressed.
For a large ratio $E_J/ E_{CJ}$, the asymptotic formula for $E_{J2}$ in the blochnium regime may be computed as the dual of the two-instanton correction to the transmon charge dispersion \cite{Zinnjustin2004,Koch2007, Dunne2017}:
\begin{equation}
    E_{J2} \sim \frac{32}{\pi} \varepsilon_\mathrm{ps} \left( \frac{1}{2} \ln \frac{\varepsilon_\mathrm{ps}}{E_L} + \ln \frac{16}{\pi} - \psi(1)\right) \exp \left(- \frac{8}{\pi} \sqrt{ \frac{\varepsilon_\mathrm{ps}}{E_L}}\right) \ ,
\end{equation}
where $\psi$ is the digamma function.
In order to remain in a soft transmon regime and keep charge dispersion under control, one would then have to lower $E_{C_S}$ to unrealistically small values, which would at any rate result in bringing down all the transition frequencies and making the qubit's operation at finite temperature impractical.
\end{document}